\documentclass[aps,prx,twocolumn,amsmath,amssymb,superscriptaddress,longbibliography,floatfix]{revtex4-2}
\usepackage{graphicx}
\usepackage{dcolumn}
\usepackage{bm}
\usepackage{hyperref}
\usepackage[mathlines]{lineno}
\usepackage{algorithm}
\usepackage[noend]{algpseudocode}
\usepackage{xcolor}
\usepackage{pifont}
\usepackage{tikz}
\usepackage[normalem]{ulem}
\usetikzlibrary{matrix}
\usepackage{mathtools}
\usepackage{fouriernc}
\usepackage[T1]{fontenc}
\usepackage{multirow}
\usepackage{physics}

\newcommand\Po{\hat{\rm P}}
\newcommand\Xo{\hat{\rm X}}

\newcommand{\Sv}{\mathbf{S}}

\newcommand{\Av}{\mathbf{A}}
\newcommand{\Gv}{\mathbf{G}}
\newcommand{\pt}{\psi_{\theta}}
\newcommand{\te}{\theta}
\newcommand{\pd}[1]{\frac{\partial \psi_{\theta}}{\partial \theta_{#1}}}
\newcommand{\pdd}[2]{\frac{\partial #1}{\partial \theta#2}}
\newcommand{\pds}[2]{\frac{\partial^2 \psi_{\theta}}{\partial \theta_{#1}\partial \theta_{#2}}}
\newcommand{\pdsrr}[3]{\frac{\partial^2 #3}{\partial \theta_{#1}\partial \theta_{#2}}}
\newcommand{\pdscr}[3]{\frac{\partial^2 #3}{\partial \theta_{#1}^*\partial \theta_{#2}}}
\newcommand{\pdsrc}[3]{\frac{\partial^2 #3}{\partial \theta_{#1}\partial \theta_{#2}^*}}
\newcommand{\pdscc}[3]{\frac{\partial^2 #3}{\partial \theta_{#1}^*\partial \theta_{#2}^*}}
\newcommand{\jac}[1]{\hat{O}_{#1}}

\newcommand{\argmin}[0]{\operatorname*{argmin}}

\newcommand\ml{\mathcal{L}}
\newcommand\mH{H}
\newcommand\Om{\Omega}

\newcommand{\up}{\uparrow}
\newcommand{\dn}{\downarrow}
\newcommand{\at}{ansatz }
\newcommand{\ats}{ans\"atze }
\newcommand{\atend}{ansatz}
\newcommand{\atsend}{ans\"atze}

\newcommand{\p}[2]{\psi_{\alpha_{#1}\beta_{#2}}}
\newcommand{\e}[2]{\epsilon^{\beta_{#1}\alpha_{#2}}}

\newcommand{\pbr}{\textit{practical limitation to reach universal representability}}

\newcommand{\eqd}{\textit{effective quantum dimension}}
\newcommand{\cutoff}{$10^{-5}$}
\usepackage{float}

\usepackage{soul}
\usepackage{changes}

\begin{document}

\title{Efficiency of neural quantum states in light of the quantum geometric tensor}

\author{Sidhartha Dash}
\affiliation{Coll\`ege de France, Universit\'e PSL, 11 place Marcelin Berthelot, 75005 Paris, France}
\author{Luca Gravina}
\affiliation{Institute of Physics, \'Ecole Polytechnique F\'ed\'erale de Lausanne (EPFL), CH-1015, Lausanne,
Switzerland}
\affiliation{Center for Quantum Science and Engineering, \'Ecole Polytechnique Fédérale de Lausanne (EPFL), CH-1015 Lausanne, Switzerland}
\author{Filippo Vicentini}
\author{Michel Ferrero}
\affiliation{CPHT, CNRS, Ecole Polytechnique, IP Paris, F-91128 Palaiseau, France.}
\affiliation{Coll\`ege de France, Universit\'e PSL, 11 place Marcelin Berthelot, 75005 Paris, France}

\author{Antoine Georges}
\affiliation{Coll\`ege de France, Universit\'e PSL, 11 place Marcelin Berthelot, 75005 Paris, France}
\affiliation{Center for Computational Quantum Physics, Flatiron Institute, New York, New York, 10010, USA.}
\affiliation{CPHT, CNRS, Ecole Polytechnique, IP Paris, F-91128 Palaiseau, France.}
\affiliation{DQMP, Universit\'e de Gen\`eve, 24 quai Ernest Ansermet, CH-1211 Gen\`eve, Switzerland.}

\date{\today}

\begin{abstract}

Neural quantum state (NQS) \ats have shown promise in variational Monte Carlo algorithms by their theoretical capability of representing any quantum state. However, the reason behind the practical improvement in their performance with an increase in the number of parameters is not fully understood. 
In this work, we systematically study the efficiency of a shallow neural network to represent the ground states in different phases of the spin-1 bilinear-biquadratic chain, as the number of parameters increases. 
We train our \at by a supervised learning procedure, minimizing the infidelity w.r.t. the exact ground state. 
We observe that the accuracy of our ansatz improves with the network width in most cases, and eventually saturates. We demonstrate that this can be explained by looking at the spectrum of the quantum geometric tensor (QGT), particularly its rank. By introducing an appropriate indicator, we establish that the QGT rank provides a useful diagnostic for the practical representation power of an NQS \atend.




\end{abstract}
\maketitle
\section{Introduction}

Recent years have seen an immense growth in the use of machine learning (ML) methods in the field of quantum many-body physics.
Central to this intersection are Neural Quantum States (NQSs), which are currently revolutionizing 
Variational Monte Carlo (VMC) approaches and related applications ~\cite{carleo2017solving,nomura2017restricted,choo2019two,hibat2020recurrent,Vicentini2019PRL,nomura2017restricted,robledo2022fermionic,lovato2022hidden,kim2023neural}.
Their success relies on the \textit{expressivity} of Neural-Networks (NNs), which have the theoretical capacity to represent any state, with a large enough number of parameters.
This is formalized by so called (i) \textit{universal representation theorems}, asserting that the approximation error inherent in a neural network (and by extension, in an NQS) can approach zero asymptotically as one increases the network's width or depth, contingent on locating the global minimum of the loss function \cite{cybenko1989approximation,leshno1993feedforward,lu2017expressive,hornik1989multilayer}.
Additionally, it is well-understood in the standard machine learning context that (ii) increasing the number of parameters beyond the \textit{over-parameterization} limit leads to a smoother loss landscape and faster convergence \cite{li2018learning,allen2019convergence}.

Building on those principles, numerous variational studies have increased the width or depth of NNs in order to check convergence in calculations where the ground state is found by energy minimization \cite{carleo2017solving,pei2021neural,viteritti2023transformer,chen2024empowering}. 
A striking example can be found in Fig.~2 of Ref.~\cite{chen2024empowering}, where the NQS approximation error at convergence systematically decreases as the number of parameters of a ResNet increases, ultimately reaching numerical precision at the largest parameter size. 
This approach is quite general and is also employed when simulating the real-time dynamics \cite{SchmittPRL2020Dynamics}, and the steady-state in open quantum systems \cite{Vicentini2019PRL,vicentini2022positive}.

However, exceptions to this rule do emerge in practical calculations, i.e. instances in which an NQS fails to become more accurate as the number of parameters 
increases. 
We will call this situation a \pbr. 
We stress that this is not in opposition to the representation theorems, which are valid only when the global minimum can be found, in the limit of a sufficiently large number of parameters. 
A \pbr~with a specific minimization algorithm might occur because the additional parameters cannot be used effectively even after the minimization has converged. As a result, the global minima cannot be found with a reasonable computational budget.

Moreover, it is unclear what the additional parameters encode once the optimization has converged. Several results in standard ML tasks~\cite{fukumizu1996regularity,karakida2019universal,XuNeurips2018Overparametrisation}, and Variational Quantum Algorithms~\cite{Larocca2023NatOverparametrisation} document the existence of redundant directions in the parameter space, suggesting that the encodings are \textit{locally} highly degenerate. 
This is in clear opposition to tensor networks and matrix product states in particular, where increasing the bond dimension is linked with the increase in the maximal entanglement entropy of the state \cite{hastings2007area,TagliacozzoPRB2008EntanglementMPS}. 
Instead, while it has been shown that even simple NQSs can encode states with arbitrary entanglement~\cite{deng2017quantum,levine2019quantum,gauvin2023mott}, 
we do not know what states NQSs with a finite number of parameters cannot encode.
The role of parameters in an NQS and their relationship to the overall accuracy of a calculation is therefore unclear. And while in some cases it is possible to invoke some theorems as an explanation for the effectiveness of increasing the number of parameters, we still do not understand what happens in presence of a \pbr.

Nevertheless, it is possible to quantify the role of parameters in an NQS by looking at the Quantum Geometric Tensor (QGT), which is a generalization of the Fisher Information Matrix (FIM) in the context of variational wavefunctions~\cite{stokes2020quantum,park2020geometry}. 
In addition to its usage in the stochastic reconfiguration (SR) method~\cite{sorella1998green,sorella2005wave}, a few studies have used the QGT to identify the relevant parameters in a variational optimization~\cite{haug2021capacity,medvidovic2023variational}.

In this work, we investigate the practical representation ability of a restricted Boltzmann Machine (RBM) as an NQS \at in a simple setting: 
the search for the ground state of a one-dimensional quantum spin model. 
We choose the spin-1 bilinear-biquadratic (BLBQ) chain because of its diverse phase diagram encompassing the topologically ordered gapped Haldane phase and the gapless extended critical phase. This makes it an ideal test bed for our study.
To directly investigate the representation ability of the NQS \atend, we choose to employ a supervised learning procedure with the NQS to represent a given ground state, i.e. we optimize the infidelity of the NQS \at w.r.t. the exact ground state. 
We perform infidelity optimizations at various network densities of the NQS \at to study the impact of network density on the NQS ground state approximation. 
We use an NQS \at given by a modified RBM (see Methods section~\ref{sec:methods}) suitable for spin-1 systems~\cite{pei2021neural}.
We optimize the infidelity using the natural gradient descent (NGD) algorithm~\footnote{Natural gradient descent (NGD)~\cite{amari1998natural,amari1998why} is essentially a generalized version of the stochastic reconfiguration (SR) method~\cite{sorella2005wave}, applicable to any supervised learning procedure. The term `stochastic reconfiguration' is specifically used when NGD is implemented with the quantum geometric tensor (QGT), for energy minimization in variational Monte Carlo (VMC) methods. Here, we implement NGD for infidelity minimizations with the QGT (see Methods section~\ref{sec:methods} for more details), instead of the Fisher Information Matrix (FIM) since we deal with complex wavefunctions in general}, where the gradients and the QGT matrix are computed exactly to exclude biases induced from sampling in the Monte Carlo procedure. 

To analyze our results, we propose a notion of the {\it dimension of the relevant manifold} $d_r$ of a variational \atend, given by the rank of the QGT (described in details in section~\ref{par:eqd}). Furthermore, we also identify a strict upper bound for the rank of the QGT, which we call the \eqd~$d_q$, given the task of representing the ground state of a particular Hamiltonian. We establish that the ratio $d_r/d_q$ can be used as a local probe to understand the evolution of the NQS approximation with the size of the network, as well as the case of a \pbr.
Furthermore, we demonstrate that $d_r/d_q$ is an important indicator for characterizing the efficiency of an NQS \at in representing a given ground state as the network width increases.

The article is structured as follows: In section II, we describe the model Hamiltonian and its phase diagram, for which we investigate NQS approximations for the different ground states. 
In section III, we describe our main results, including the evolution of the NQS approximation with the width of the network, and the role of the quantum geometric tensor (QGT) as a diagnostic tool. 
In section IV, we interpret these results, emphasizing the QGT's role in understanding the success and limitations of an NQS, and summarize the main findings.


\section{The bilinear biquadratic spin-1 chain}

\begin{figure}[h!]   
\centering
\includegraphics[scale=0.5]{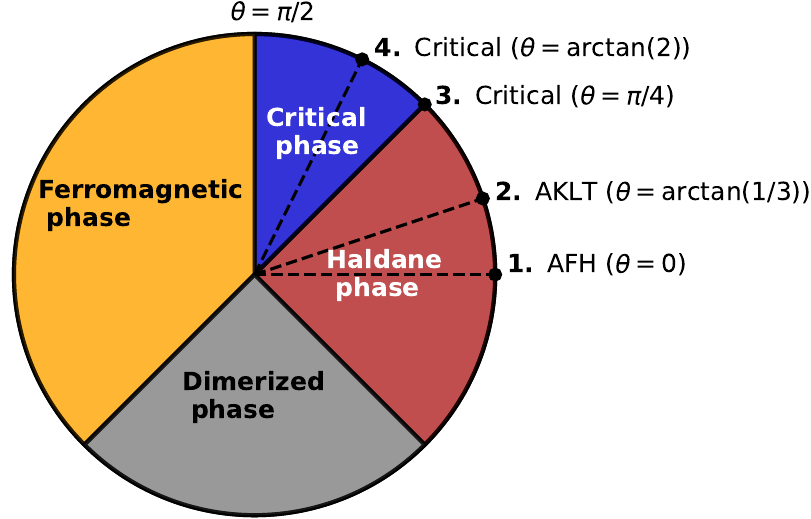}
\caption{Phase diagram of the spin-1 bilinear-biquadratic (BLBQ) chain.
We perform infidelity (Eq.~\eqref{eq:infidelity}) minimization,
with a Neural Quantum State (NQS) \at given by a modified Restricted Boltzmann Machine (RBM) for spin-1 models~\cite{pei2021neural} (described in Methods section ~\ref{sec:methods}), at the four marked points in the phase diagram.}
\label{fig:fig_cartoon}
\end{figure}

{\it Model}. 
The spin-1 bilinear-biquadratic (BLBQ) model in one dimension is defined by the Hamiltonian~\footnote{More generally, the Hamiltonian is written as $\displaystyle\mH = \sum_i J \left[\cos(\theta)\Sv_i\cdot \Sv_{i+1} + \sin(\theta)\left(\Sv_i\cdot \Sv_{i+1}\right)^2\right]$.}:
\begin{align}
   \label{eq:BLBQ} 
    \mH &= \sum_i J \left[\Sv_i\cdot \Sv_{i+1} + \tan(\theta)\left(\Sv_i\cdot \Sv_{i+1}\right)^2\right].
\end{align}
The model is parametrized by an angular variable $\theta$, and $\Sv_i=(S_{i}^x,S_{i}^y,S_{i}^z)$ is the spin operator acting on the local spin-1 Hilbert space at site $i$.
The above Hamiltonian has a gapped Haldane phase ~\cite{haldane1983continuum,den1989preroughening} 
for $-\pi/4<\theta<\pi/4$, an extended critical phase~\cite{lauchli2006spin,uimin1970one,lai1974lattice,sutherland1975model,itoi1997extended} for $\pi/4 \leq \theta<\pi/2$, a ferromagnetic phase for $\pi/2\leq\theta <5\pi/4$, and a dimerized phase for $-3\pi/4<\theta<-\pi/4$~\cite{parkinson1988s,barber1989spectrum} (Fig.~\ref{fig:fig_cartoon}).
We focus on the Haldane phase and the extended critical phase in this work.
The Haldane phase describes the isotropic Heisenberg antiferromagnet at $\theta=0$.
It is characterized by a hidden topological order, which is the strongest for the {\it Afﬂeck-Kennedy-Lieb-Tasaki} (AKLT) state~\cite{affleck1988valence,schollwock1996onset,affleck2004rigorous}  at $\theta=\arctan(1/3)$.
The AKLT state is a valence-bond state, which can be represented by two spin-1/2 particles at each site, forming singlets with the spins of the neighboring sites (see supplementary material sec.~\ref{sec:AKLT}).
The AKLT state has an exact matrix product state (MPS) representation with bond dimension 2~\cite{schollwock2011density}. It is also the state that has the lowest bipartite entanglement in comparison to ground states at other values of $\theta$~\cite{dai2022commensurate}. 
The Haldane phase is gapped and has a ground state with an exponentially decaying antiferromagnetic spin-spin correlation 
in the thermodynamic limit. Up to the AKLT point, the correlation function behaves as 
$\langle S^{\alpha}_0\cdot S^{\alpha}_j\rangle\approx (-1)^j \exp(-j/\xi)/\sqrt{j}$
with a correlation length $\xi=1/\ln(3)$ for the AKLT state. 
For $\arctan(1/3)<\theta<\pi/4$, the modulation wave-vector shifts away from $k=\pi$ to reach 
$k=2\pi/3$ at $\theta=\pi/4$. The Haldane gap closes at $\theta=\pi/4$, marking a critical point, known as the {\it Uimin-Lai-Sutherland} (ULS) point~\cite{uimin197412,lai1974lattice,sutherland1975model}, across a {\it Berezinskii-Kosterlitz-Thouless} (BKT) transition~\cite{itoi1997extended,binder2020low}. At this point, the model  has an exact $SU(3)$ symmetry and is integrable. Its low-energy physics is described by the Wess-Zumino-Witten ($\rm SU(3)_{\rm k=1}$) conformal field theory~\cite{schollwock1996onset,itoi1997extended,binder2020low}. The model remains gapless in the thermodynamic limit throughout the region $\pi/4 \leq \theta<\pi/2$. 
In this phase, the dominant correlations are of quadrupolar nature, with wavevector $k=\pm 2\pi/3$, and decay as a power law with exponent $\eta=4/3$~\cite{itoi1997extended,lauchli2006spin}.
\begin{figure*}[ht]
\centering
\includegraphics[width=\textwidth]{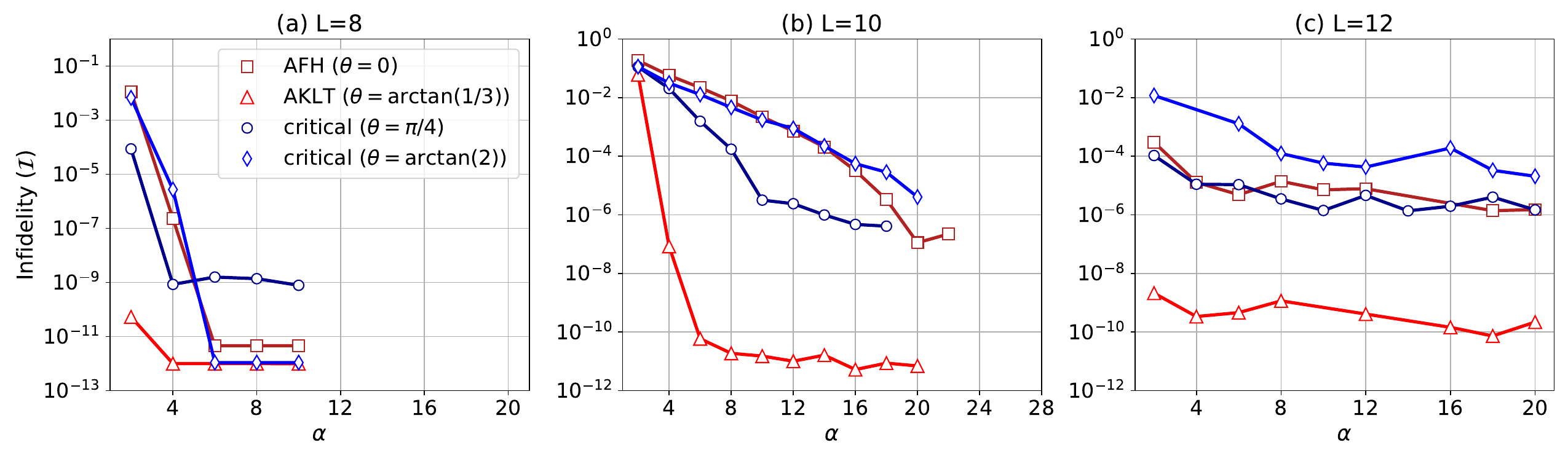}
\caption{Infidelity and the hidden layer density $\alpha$ for infidelity minimization on the spin-1 BLBQ chain. Infidelity ($I$) after the convergence of an exact infidelity minimization procedure is shown as a function of $\alpha$, for the AFH, AKLT and the critical phases (corresponding to the four marked points in Fig.~\ref{fig:fig_cartoon}) of the spin-1 BLBQ chain with open boundary conditions. Results are shown for chain lengths (a) $L=8$, (b) $L=10$, and (c) $L=12$. The NQS used is a modified RBM for spin-1 systems~\cite{pei2021neural} (see Methods section~\ref{sec:methods}).}
\label{fig:fig_infid}
\end{figure*}

\begin{figure*}
\centering
\includegraphics[width=0.95\textwidth]{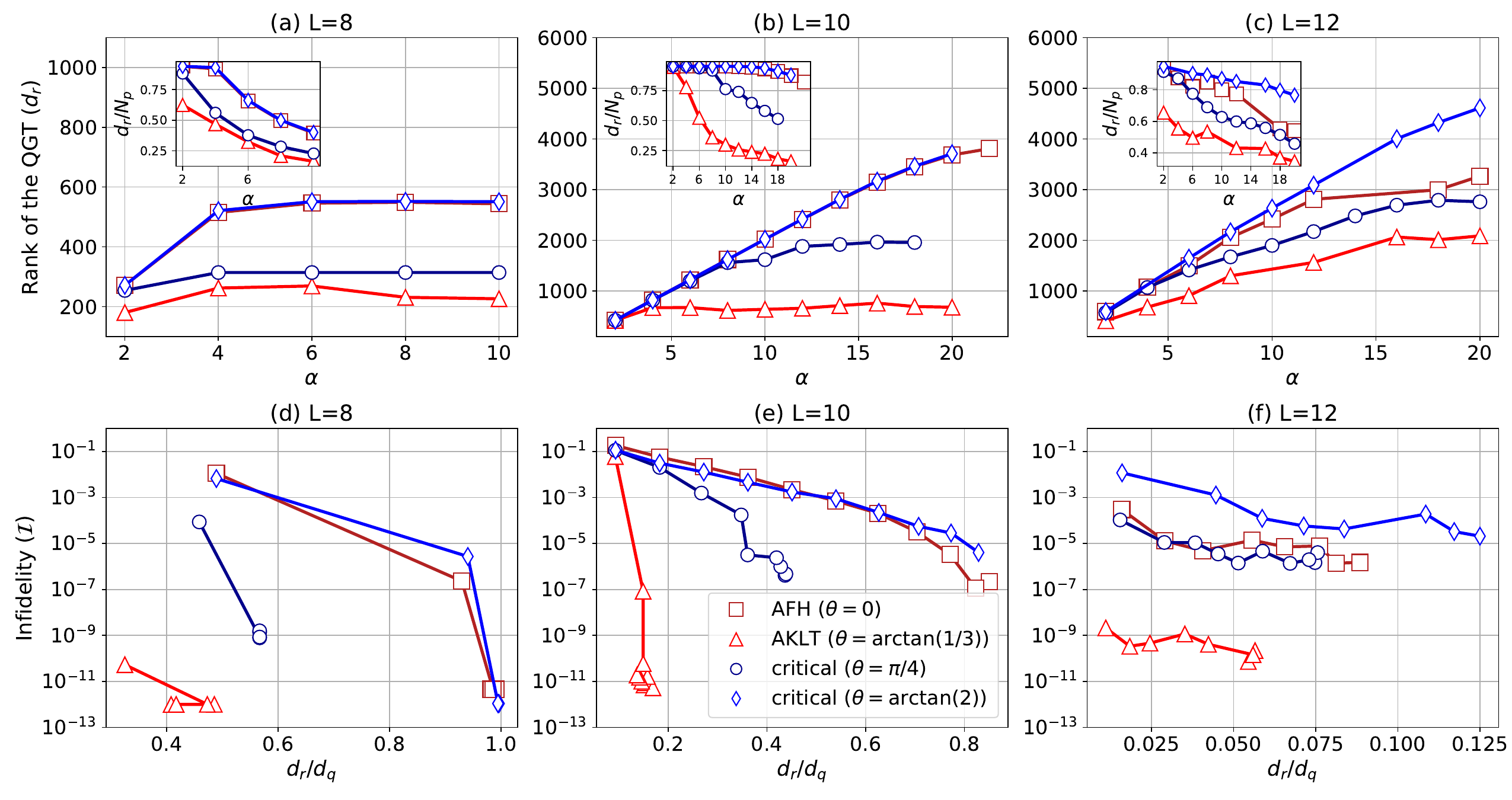}
\caption{QGT rank and converged infidelity for infidelity minimizations on the spin-1 BLBQ chain. (a)-(c) The rank of the QGT ($d_r$) is shown as a function of the hidden layer density $\alpha$ for lengths $L=8,\;10,\;{\rm and}\;12$. The insets show the QGT rank normalized w.r.t. the number of parameters $N_p$. (d)-(f) The converged infidelity ($I$) is shown as a function of the ratio $d_r/d_q$,
where $d_q$ is the upper bound for the QGT rank (or the \eqd), with $d_q = 554,\;4477,\;{\rm and}\;36895$ respectively for $L=8,\;10,\;{\rm and}\;12$. The QGT rank $d_r$ (which is also the {\it dimension of the relevant manifold} around the converged solution) is computed as the number of eigenvalues of the QGT greater than \cutoff. Each panel shows the data for the AFH, AKLT, and the two critical phases of the spin-1 BLBQ chain with open boundary conditions, for a given length of the chain.}
\label{fig:fig_qgt_rank}
\end{figure*}
\begin{figure*}
\centering
\includegraphics[width=0.9\textwidth]{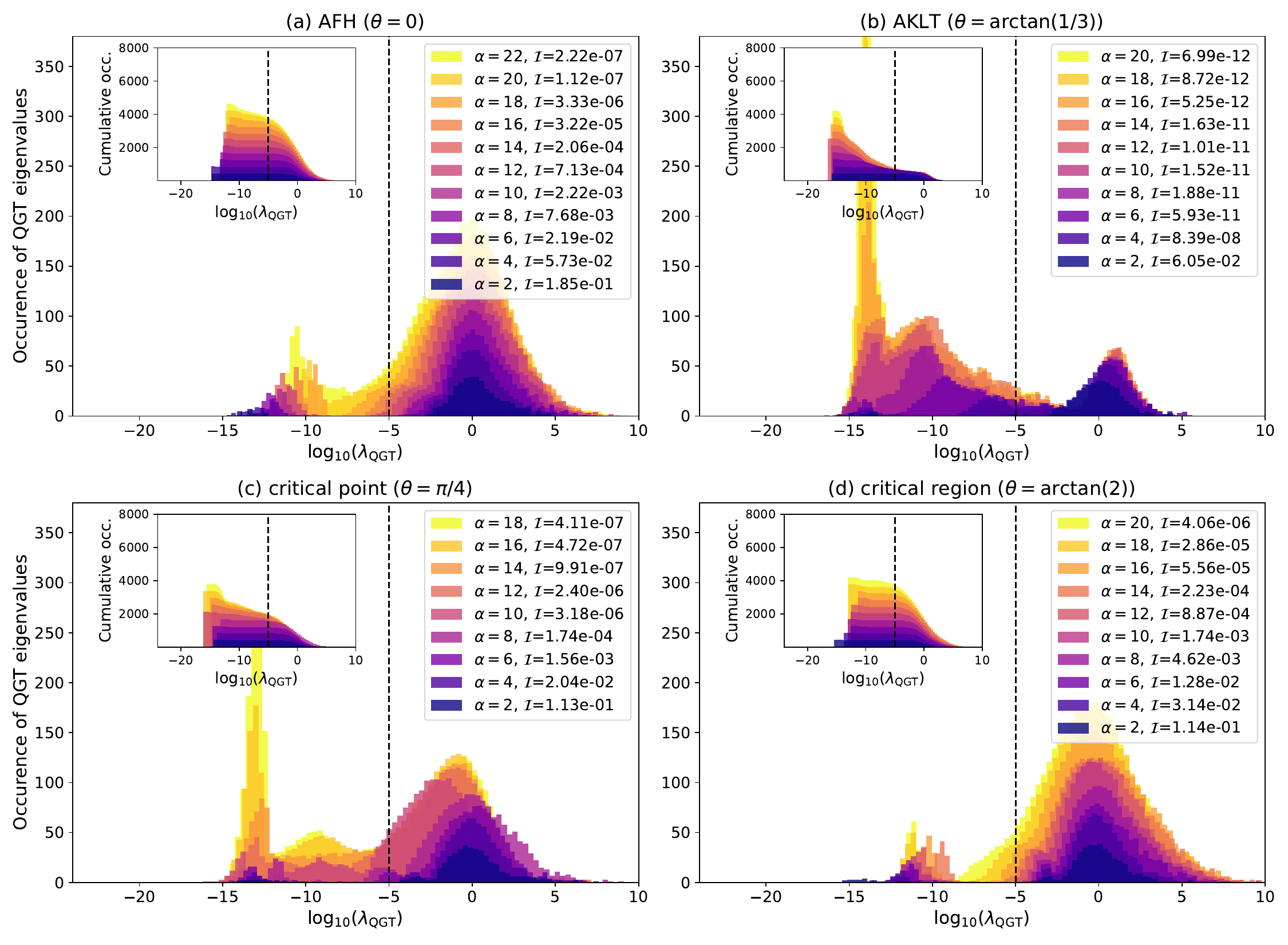}      
\caption{QGT spectra for infidelity minimization on the spin-1 BLBQ chain. The spectra of the QGT is shown for the four phases (a)-(d) (corresponding to the four marked points in Fig.~\ref{fig:fig_cartoon}) after convergence of the infidelity minimization of the NQS w.r.t. the exact ground state of the spin-1 BLBQ chain with length $L=10$. The QGT eigenvalues are shown in $\log_{10}$ scale. The insets show the cumulative distribution of the eigenvalues (number of eigenvalues exceeding the value on the $x$-axis) for each case. The dashed line marks the cutoff \cutoff~, used for computing the rank $d_r$.}
\label{fig:fig_qgt}
\end{figure*}    

\section{Results}

We search for NQS approximations of the ground states in the Haldane and the critical phases of the spin-1 BLBQ chain by minimizing the infidelity of the NQS w.r.t. the exact ground state $\ket{\Om}$ (see Methods section ~\ref{sec:methods} for more details),
\begin{align}
\label{eq:infidelity}    \textit{I} = 1-\frac{\braket{\pt}{\Om}\braket{\Om}{\pt}}{\braket{\pt}{\pt}\braket{\Om}{\Om}}.
\end{align}
Here $\ket{\pt}=\sum_{n}\pt(n)\ket{n}$ is the NQS and  
$\ket{n}$ denotes a state in the computational basis.  
Note that the degree to which the NQS \at represents the true ground state is given by the fidelity
$\bm{f} = 1-I \in [0,1]$, and hence the infidelity $I$ gives the error in the NQS approximation for the ground state. 
For comparisons, we also perform the usual VMC procedure, included in the supplementary material sec.~\ref{sec:qgt_spectrum_vmc}, where one minimizes the energy \begin{align}
\label{eq:energy}    E = \langle H \rangle  = \sum_n p_{\theta}(n)\frac{\bra{n}\mH\ket{\pt}}{\braket{n}{\pt}},
\end{align}
to get an NQS approximation for the ground state. Here $p_{\theta}(n)=|\pt(n)|^2/\braket{\pt}{\pt}$ is the probability distribution over the computational basis. 
For the rest of the paper, we focus on the results of the infidelity minimization procedure. We restrict our computations to the total $S_z=0$ sector of the Hilbert space. 
 

We report the infidelities at convergence of the infidelity minimization with NGD in Fig.~\ref{fig:fig_infid}, for different sizes of the BLBQ chain $L=8,\;10,\;12$, for four phases in the phase diagram (the four points marked in Fig.~\ref{fig:fig_cartoon}), and for different hidden layer densities $\alpha$. 
As we increase $\alpha$, the converged infidelities for the different phases reduce to very low values and saturate thereafter at a certain value of $\alpha$ in most cases for $L=8,\;10$. 
The converged infidelities for the four phases for $L=12$ oscillate after an initial descent as $\alpha$ increases. 
We also remark that the NQS for the AKLT point has the best approximation for the ground state (the minimum infidelity) in most cases.
This is consistent with the fact that the AKLT state has the lowest entanglement (has an exact MPS representation with bond dimension 2), among all the other phases. 


To avoid artefacts originating from the Monte Carlo sampling~\cite{sinibaldi2023unbiasing} and solely investigate the representation power of the NQS ansatz, we perform the infidelity minimization by summing over the total $S_z=0$ subspace of the Hilbert space. 


Additionally, we check that the infidelity minimization has converged well by inspecting the local infidelity landscape through the Hessian (see supplementary material~\ref{sec:hessian_spectra} for the spectra of the Hessian for $L=8$). Our solutions lie in a deep valley (many large positive eigenvalues), which has a few infinitesimally small downward slopes (few very small negative eigenvalues) as commonly seen in the ML literature~\cite{dauphin2014identifying,fort2019goldilocks}.

A systematic decrease in the NQS approximation error with $\alpha$ to numerical precision, followed by a saturation, is consistent with our expectations from the \textit{universal representation theorems}. 
This is the case for infidelity optimizations for the AFH, AKLT and the critical phase at $\theta=\arctan(2)$ with $L=8$, as well as for the AKLT phase with $L=10$, where we achieve converged 
infidelities $\leq10^{-11}$. For practical purposes, we consider having reached \textit{numerical precision} in the above cases. 
For the other cases, the infidelities either saturate at a larger value with small oscillations, which we call a \textit{premature saturation} of the infidelity, or continue to decrease with $\alpha$. The latter could suggest that the NQS needs more parameters to be able to approximate the solution better. However, a \textit{premature saturation} of the infidelity suggests that the optimized NQS is locally not able to use the newly added parameters to improve the solution. This scenario can be further investigated by examining the Fubini-Study (FS) metric in the space of variational wavefunctions. From this metric, we derive the quantum geometric tensor (QGT)~\cite{park2020geometry,stokes2020quantum}, which gives a measure of the \textit{relevant directions} in the parameter space, as discussed in the following paragraphs.


\paragraph*{Quantum Geometric Tensor (QGT)}
In defining a distance measure between two pure states, one must consider that states differing by a global phase are physically indistinguishable \cite{patel2024natural}. Consequently, the set of pure states is isomorphic to the set of rays in the Hilbert space, i.e. to the projective Hilbert space, over which a metric can be unambiguously defined \cite{brody2001geometric, facchi2010classical}. 
The Fubini-Study (FS) metric naturally emerges as the unique (up to a constant factor) Riemannian metric on the set of rays, being invariant under all unitary transformations, that is, under all possible time evolutions \cite{wootters1981statistical}.
For two pure states $\ket{\psi_\theta}$ and $\ket{\psi_\phi}$, parameterized by $\theta$ and $\phi$, the FS distance is defined as:
\begin{equation}
\text{FS}(\psi_\theta, \psi_\phi) = \arccos\left(\sqrt{\frac{\langle \psi_\theta | \psi_\phi \rangle \langle \psi_\phi | \psi_\theta \rangle}{\langle \psi_\theta | \psi_\theta \rangle \langle \psi_\phi | \psi_\phi \rangle}}\right).
\end{equation}

For an infinitesimal variation $\phi = \theta + d\theta$, a second-order expansion of the squared FS distance yields the line element:
\begin{equation}
ds^2 = \sum_{\alpha,\beta} {\rm G}_{\alpha\beta}d\theta_{\alpha}^*d\theta_{\beta},
\end{equation}
where the QGT~\footnote{This expression holds for the case when the variational state $\ket{\pt}$ is normalized. In the general case, the QGT takes the form ${\rm G}_{ij} = \frac{\braket{\pd{i}}{\pd{j}}}{\braket{\pt}{\pt}} - \frac{\braket{\pd{i}}{\pt}\braket{\pt}{\pd{j}}}{\braket{\pt}{\pt}^2}$.}
\begin{align}
\label{eq:QGT}
{\rm G}_{\alpha\beta}
=\braket{\frac{\partial\psi_{\theta}}{\partial\theta_{\alpha}}}{\frac{\partial\psi_{\theta}}{\partial\theta_{\beta}}} - \braket{\frac{\partial\psi_{\theta}}{\partial\theta_{\alpha}}}{\psi_{\theta}}\braket{\psi_{\theta}}{\frac{\partial\psi_{\theta}}{\partial\theta_{\beta}}}
\end{align}
captures the local curvature of the space.
This matrix is positive semi-definite, ensuring that its eigenvalues are real and non-negative. 
These eigenvalues measure the curvature of the FS distance along the associated eigendirections. Non-zero eigenvalues correspond to \textit{relevant directions} where the quantum state undergoes significant variation, while zero eigenvalues indicate \textit{redundant directions} where parameter changes produce no meaningful change in the quantum state \cite{park2020geometry}.
Thus, at convergence, the rank of the QGT reflects the number of \textit{relevant directions} ($d_r$), or the \textit{dimension of the relevant manifold}, that the variational ansatz uses to represent the quantum state.
Notably, in the special case where wavefunction amplitudes are real and positive, that is $\ket{\psi_\theta} = \sum_n \sqrt{p_\theta(n)} \ket{n}$, the QGT simplifies to a scaled version of the classical Fisher Information Matrix (FIM) \cite{stokes2020quantum}. Specifically,
\begin{align}
\label{eq:FIM-QGT}
{\rm G}_{\alpha\beta}=&\frac{1}{4}\left\langle\frac{\partial\log p_{\theta}(n)}{\partial \theta_{\alpha}}\frac{\partial\log p_{\theta}(n)}{\partial \theta_{\beta}}\right\rangle \nonumber\\
&-\frac{1}{4}\left\langle\frac{\partial\log p_{\theta}(n)}{\partial \theta_{\alpha}}\right\rangle\left\langle\frac{\partial\log p_{\theta}(n)}{\partial \theta_{\beta}}\right\rangle = \frac{1}{4}\mathcal{F}_{\alpha\beta},
\end{align}  
where the averages are taken with respect to the Born probability distribution $p_{\theta}(n)$ over the states $\ket{n}$. $\mathcal{F}$ is the FIM of the probability distribution $p_{\theta}(n)$ and arises from the second-order expansion of the Kullback-Leibler divergence, a proximity measure between probability distributions.

\paragraph*{Natural Gradient Descent (NGD).}
Standard gradient descent (GD) algorithms attempt to
minimize the loss function iteratively by moving in the direction of steepest descent 
under Euclidean geometry. This is done following the prescription 
\begin{equation}
  \delta\theta = - \eta\;\bm{F},  
\end{equation}
where $\delta\theta$ is the parameter update, $\eta$ is the learning rate and $\bm{F}$ is the conjugate gradient of the loss function in Eq.~\eqref{eq:infidelity} (see Methods section ~\ref{sec:methods}). 
As a result of assuming an underlying Euclidean geometry, GD is poorly suited to navigate the projective Hilbert space
due to the geometry of the space 
itself, 
and 
the parameter redundancy introduced by a neural network parametrization \cite{patel2024natural}. Thus standard GD often struggles in landscapes with steep valleys or flat plateaus where convergence can be slow as updates 
oscillate in steep directions while taking small steps in shallow ones.

NGD improves upon this by adapting parameter updates to the intrinsic geometry of the function space, as defined by the QGT \cite{amari1998natural,amari1998why,gravina2024neural, stokes2020quantum}. Instead of relying on the Euclidean gradient, NGD employs the geometry-aware update rule (see supplementary material ~\ref{sec:opt-details}):
\begin{equation}
\delta\theta = -\eta \; \Gv^{-1} \bm F.
\end{equation}
The action of the QGT inverse $\Gv^{-1}$ can be heuristically understood as locally ``flattening'' the landscape. 
NGD offers several key advantages. First, by adjusting for local curvature, NGD prevents overshooting in steep directions and accelerates progress in flat regions. Furthermore, NGD is first-order invariant under reparameterization, ensuring consistent optimization across different representations. Ultimately, NGD is less prone to slow convergence near saddle points or local minima guaranteeing faster convergence under a variety of conditions \cite{martens2020new, zhang2019fast}.

\paragraph*{Effective quantum dimension.} 
\label{par:eqd}
One can express Eq.~\eqref{eq:QGT} as 
\begin{align}
\label{eq:QGT-mat}
{\rm G}_{\alpha\beta} = \left\langle(\jac{\alpha}^{\dagger}-\langle\jac{\alpha}^{\dagger}\rangle)(\jac{\beta}-\langle \jac{\beta} \rangle) \right\rangle,    
\end{align}
where $\jac{\mu} = \sum_n \partial/\partial \theta_{\mu} \ket{n}\bra{n}$ is a partial derivative operator, and $\ket{n}$ is a state in the computational basis. Eq.~\eqref{eq:QGT-mat} can then be represented as a matrix multiplication $\sum_n M^{\dagger}_{\alpha n}M_{n \beta}$, where the matrix elements $M_{n\mu} = \bra{n}(\jac{\mu}-\langle \jac{\mu}\rangle)\ket{\pt}$, and the summation over $n$ runs over the relevant (possibly constrained by symmetries) Hilbert space. We introduce the notion of an \eqd~$d_q$, which represents the dimension of this space. This gives the upper bound on the QGT rank $d_r$, given the task of representing the ground state of a specific Hamiltonian. In our case, for the spin-1 chain with the constraint total $S_z=0$, and a global spin-flip symmetry, the \eqd~$d_q=(D+1)/2$, where $D$ is the dimension of the total $S_z=0$ subspace. Therefore, $d_q=(D+1)/2$ is the maximum \textit{dimension of the relevant manifold} around the optimized NQS. For our computations, $d_q$ takes the values $554,\;4477$, and $36895$ respectively for $L=8,\;10$, and $12$ (for all four phases). 


\paragraph*{Rank of the QGT.} 
We compute the QGT at convergence of the infidelity minimization procedure for each case, and plot the rank of the QGT ($d_r$) with the hidden layer density $\alpha$ in Fig.~\ref{fig:fig_qgt_rank}(a)-(c). The insets show the QGT rank normalized w.r.t. the number of parameters ($N_p$) in the NQS \atend. The QGT rank $d_r$ increases with an initial increase in $\alpha$, accompanied by a decrease in the infidelity. This suggests a correlation between the accuracy of the NQS approximation and $d_r$. To further elucidate this correlation, we plot the converged infidelity as a function of the ratio of the QGT rank w.r.t. the \eqd, i.e. $d_r/d_q$, in Fig.~\ref{fig:fig_qgt_rank}(d)-(f). This reveals that the infidelity decreases exponentially with the ratio $d_r/d_q$ until it saturates beyond a particular value of $\alpha$. 
For the AFH phase and the critical phase at $\theta=\arctan(2)$, with L=8, we achieve \textit{numerical precision} at the saturation of the ratio $d_r/d_q\approx 1$ (when the QGT rank takes its maximum value), i.e. when the NQS is performing at its best. While, for the AKLT phases with $L=8$ and $10$, we attain \textit{numerical precision} even with $d_r/d_q<1$, when the NQS is not yet at its full capacity. Furthermore, the NQS also performs very well for the AKLT phase with $L=12$ despite a significantly lower $d_r/d_q$ compared to the other phases. This implies that the wavefunction for the AKLT phase is inherently simpler to be represented by the NQS. Whereas, in cases where the infidelity saturates at higher values with some oscillations, 
the ratio $d_r/d_q$ plateaus below $1$. This suggests that the NQS is not able to add \textit{relevant directions} around the converged solution, with an increase in $\alpha$, which in turn limits its ability to improve its approximation. 
Note, however, that we have a small deviation from this behaviour for some cases with $L=12$, where the QGT rank $d_r$ (and the ratio $d_r/d_q$) increases while the infidelity approaches \textit{premature saturation}. This could be attributed to the decreasing normalized rank $d_r/N_p$ in the inset of Fig.~\ref{fig:fig_qgt_rank}(c), suggesting that the QGT rank increases at a slower rate than the number of parameters $N_p$, leading to the inability of the optimized NQS to represent the ground states with a higher accuracy. This is in contrast to the case of $L=10$ (notably for the AFH and the critical phases), where the ratio $d_r/N_p$ remains constant in the regime where the optimized NQS consistently improves its approximation with an increase in $\alpha$. 

Moreover, for a given size of the chain, the slope of the $I \;{\rm vs.}\; d_r/d_q$ plot provides a measure of the efficiency of the NQS \at to represent a given ground state with increasing accuracy as its width grows. For instance, for $L=10$, the $I \;{\rm vs.}\; d_r/d_q$ plot has the steepest slope for the AKLT phase, suggesting that the optimized NQS \at represents the AKLT phase most efficiently. In contrast, the NQS \at represents the AFH and the critical ($\theta=\arctan(2)$) phases with a much lower efficiency.


\paragraph*{Spectrum of the QGT.}
While we demonstrated that the improvements and limitations in the NQS approximation can be analyzed using the rank of the QGT, it is also insightful to examine the spectrum of the QGT which reveals how the local metric of the FS distance, around the converged solution, evolves with $\alpha$. We plot the spectrum of the QGT, for each solution with $L=10$, as histograms in Fig.~\ref{fig:fig_qgt}. We broadly notice two regimes in the evolution of the QGT spectra as $\alpha$ increases, in relation to the converged infidelities: i) a decreasing infidelity regime (for small values of $\alpha$), and ii) a saturated infidelity regime (for relatively larger values of $\alpha$). In regime (i), the QGT spectra indicate a significant increase in large-magnitude eigenvalues as $\alpha$ increases, resulting in an increase in the QGT rank (and the ratio $d_r/d_q$, indicating a growth in the \textit{dimension of the relevant manifold} around the converged solution), as seen in Figs.~\ref{fig:fig_qgt_rank},\ref{fig:fig_qgt}. This is accompanied by a decrease in the converged infidelities. Whereas in regime (ii), the QGT spectra show an increase in very small eigenvalues as $\alpha$ increases, and hence the \textit{dimension of the relevant manifold} (and the ratio $d_r/d_q$) does not change anymore. This leads to the narrow peaks at very low eigenvalues (corresponding to the \textit{redundant directions}), in the histograms  (Fig.~\ref{fig:fig_qgt}), as can be seen clearly for the AKLT and the critical phase at $\theta=\pi/4$ with $L=10$. In this regime, the NQS approximation for the ground state no longer improves. Note that this can happen either in the scenario where we have achieved \textit{numerical precision}, or in the case of a \textit{premature saturation} of the converged infidelity. 


\section{Discussion and Conclusion}

In summary, we have performed a large number of numerical experiments to find ground states in various phases of the spin-1 BLBQ chain using a 1-layer NQS \at (Eq.~\eqref{eq:spin-1_RBM}).
We perform a supervised learning procedure, where we minimize the infidelity of the NQS w.r.t. the exact ground state (Eq.~\eqref{eq:infidelity}). By increasing the hidden layer density $\alpha$ of the network, we achieve NQS approximations with greater accuracy, sometimes even reaching \textit{numerical precision} ($I\leq 10^{-11}$). The increase in the NQS accuracy is accompanied by an increasing QGT rank (and the related quantity $d_r/d_q$). In contrast, we sometimes observe a \textit{premature saturation} in the NQS accuracy, where the NQS infidelity does not decrease further with an increase in $\alpha$. In this regime, the QGT rank (or $d_r/d_q$) saturates around a given value or grows very slowly, for NQS optimizations with different $\alpha$. This provides a strong evidence that the local representation power of an NQS ansatz can be characterized using the QGT rank (or the ratio $d_r/d_q$) at convergence, within a given optimization scheme. 



We interpret the above observations as follows. 
The rank of the QGT measures the number of directions in the Hilbert space that can be accessed by an infinitesimal change of the parameters. As a result, the QGT rank $d_r$ can be interpreted as the dimension of the tangent manifold around the optimized NQS. Therefore a higher $d_r$ suggests that the NQS \at had more freedom to reach a higher accuracy, given a particular optimization scheme. Whereas a saturation in $d_r$, with an increase in $\alpha$, limits the ability of the NQS to leverage the additional parameters for improving its accuracy, and hence leads to a \textit{premature saturation} of the NQS accuracy. We identify this as a \pbr~for the given NQS optimization. In such a case, the QGT spectrum mostly shows an increase in zero eigenvalues, corresponding to the \textit{redundant directions} around the optimized NQS \at in the parameter space. It is important to note that our results do not suggest a breakdown of the universal representation theorems, and rather demonstrate situations where practical limitations in the NQS optimization limit the accuracy of the NQS approximation with an increasing width of the NN.

Such observation of \textit{redundant directions} is consistent with established findings in the ML literature, which show that a substantial number of directions in the parameter space are redundant~\cite{fukumizu1996regularity,karakida2019universal}. This is in contrast to tensor network \atsend, where each variational parameter contributes to the entanglement entropy of the variational \atend~\cite{TagliacozzoPRB2008EntanglementMPS}, systematically improving the approximation for the ground state. Nevertheless, our analysis of the QGT rank/spectrum can help diagnose a \pbr.
When this happens, in most cases a \textit{premature saturation} of the converged infidelity is accompanied by a saturation of the QGT rank (or $d_r/d_q$), with an increasing width of the NQS \atend. 
In a few cases for $L=12$, we observe that $d_r/d_q$ does not fully saturate when the infidelity does, but rather grows at a slower rate than the number of parameters in the NQS \atend. In the former case, the optimized NQS only adds \textit{redundant directions} in its neighborhood as its width increases. 
While in the latter case, the optimized NQS \at is very less sensitive to an increasing \textit{dimension of the relevant manifold}, suggesting that the newly added directions only marginally improve the NQS approximation.  
A \textit{premature saturation} of the converged infidelity with a growing network width can arise from two independent factors: i) the limited expressivity of the specific NQS architecture to represent a given ground state, or ii) the inability of the optimization method to reach the global minimum of the cost function, or from a combination of the two. Although our analysis using the QGT rank can quantify the practical efficiency of a converged NQS \atend~and can identify when a \pbr~happens within a given optimization scheme, it cannot precisely determine whether the \textit{premature saturation} occurs due to the factor i) or ii). 
A saturation of the QGT rank at $d_r<d_q$ (as commonly observed in cases of \textit{premature saturation} of the converged infidelity) can be caused either by using a badly configured \at with redundant parameters, or by an inefficient optimization scheme.   

Nevertheless, a compelling application of our analysis lies in characterizing different methods of encoding the input for a given NQS architecture. This involves an inspection of the rank of the QGT at initialization for different encodings, and choosing the encoding with the highest rank of the QGT, which would ensure that the chosen \at is the easiest to optimize. 

To summarize the performance of the NQS \at across the four phases of the spin-1 BLBQ chain, we note that the NQS represents the ground states in the AKLT phase and the critical point at $\theta=\pi/4$ with a lower value of $d_r/d_q$ than that in the AFH phase and the critical phase at $\theta=\arctan(2)$, for a given accuracy. The NQS approximation for the AKLT state performs the best, exhibiting the lowest QGT rank $d_r$, suggesting that it is the simplest state to be represented by the given NQS \atend. Furthermore, the AKLT state is also the state with the simplest non-trivial entanglement structure, since it can be represented exactly as an MPS with bond dimension $\chi = 2$~\cite{schollwock2011density}. The NQS approximation for the critical phase at $\theta=\arctan(2)$ has the largest QGT rank $d_r$. Moreover, it is interesting to note that the NQS approximation for the ULS critical point at $\theta=\pi/4$ has a lower $d_r/d_q$ than the AFH state despite having a higher bipartite entanglement entropy~\cite{dai2022commensurate}. This may be due to the fact that the Hamiltonian at the ULS point has an enhanced SU(3) symmetry~\cite{lauchli2006spin,binder2020low}, which requires further investigation. 

Furthermore, an interesting question is how the converged infidelity, and the QGT rank vary with network width under different optimization schemes. To investigate this, we also conducted the infidelity minimization for the AFH and the AKLT phases with another algorithm: a combination of ADAM~\cite{kingma2014adam} and YOGI~\cite{zaheer2018adaptive} optimization (see section ~\ref{sec:comp_optimize} in the supplementary material), which are adaptive gradient-based optimizers that usually perform well for stochastic gradient descent (SGD). We found that the NQS approximations with ADAM+YOGI optimization were generally less accurate at large network widths than those obtained using NGD (see Fig.~\ref{sfig:fig_qgt_rank_comparison}(d)-(f)). In most cases, the ADAM+YOGI optimization led to minimal improvement of the NQS approximation with increasing network width, which appeared nearly frozen unlike in the case of NGD optimizations. Interestingly, the behaviour of the QGT rank with $\alpha$ was similar across the two optimization schemes, with $d_r$ eventually saturating with $\alpha$ (see Fig.~\ref{sfig:fig_qgt_rank_comparison}(a)-(c)). In summary, the NGD optimizations proved to be significantly more effective in finding accurate NQS approximations. 

In conclusion, we have introduced the notion of the {\it dimension of the relevant manifold} around a converged NQS, defined as the rank of the QGT $d_r$, and have proposed the ratio $d_r/d_q$ as an indicator to characterize the local representation power of an NQS \at within a given optimization scheme. We have also demonstrated the emergence of a \pbr, using the above indicator, which suggests that practical calculations with NQS-based \ats can differ considerably from the asymptotic behaviour predicted by universal representation theorems. 
However, it is important to note that our analysis holds when the QGT can be estimated exactly. In practical VMC computations, the empirical estimate of the QGT, or the neural tangent kernel (NTK)~\cite{chen2024empowering} for NQS \ats with a large number of parameters, are usually biased when sampled using wavefunction amplitudes~\cite{sinibaldi2023unbiasing}. As a result, the rank of the empirically estimated QGT/NTK does not give a reliable measure of the {\it dimension of the relevant manifold}. We leave for a future work to explore the relation between an empirically estimated QGT/NTK and the {\it dimension of the relevant manifold} around a converged solution.

\section*{Acknowledgements}
We are grateful to Giuseppe Carleo, Juan Carrasquilla, Michele Casula, Stephen Clark, Anna Dawid, Matija Medvidovi{\'c}, Fr\'ed\'eric Mila, Schuyler Moss, Javier Robledo Moreno, Christopher Roth, Andr\'e-Marie Tremblay and Agnes Valenti for useful discussions. S.D. would like to thank Olivier Simard for a careful reading of the manuscript. 
F.V. acknowledges support by the French Agence Nationale de la Recherche through the NDQM project, grant ANR-23-CE30-0018.
This work was granted access to the HPC resources of TGCC and IDRIS under the allocations A0150510609 and AD010514908 attributed by GENCI 
(Grand Equipement National de Calcul Intensif). 
We also acknowledge the use of computing resources at the Flatiron Institute, a division of the Simons Foundation. 



{
\section{Methods}
\label{sec:methods}
We use a modified version of the RBM, adapted for spin-1 systems~\cite{pei2021neural}, as the NQS ansatz $\pt(\sigma) = \sum_{h}\exp\left[\mathcal{E}(\sigma,h)\right]$, where
\begin{align}
    \label{eq:spin-1_RBM}\mathcal{E}(\sigma, h) &= \sum_{i=1}^{L}a_i\sigma_i + \sum_{i=1}^{L}A_i\sigma_i^2 + \sum_{i=1}^{M}\sum_{j=1}^{L}w_{ij}h_i\sigma _j\nonumber\\
&+\sum_{i=1}^{M}\sum_{j=1}^{L}W_{ij}h_i\sigma_j^2 + \sum_{i=1}^{M}h_ib_i
\end{align}
is the energy function of the spin-1 RBM.
Here $\theta=\{\{a_i\},\{A_i\},\{b_i\},\{w_{ij}\},\{W_{ij}\}\}$ is the set of all $2L+M+2ML$ complex parameters of the spin-1 RBM~\footnote{We remark that this definition of the NQS ansatz intrinsically uses an activation function $\log(\cosh(\cdot))$, which has branch cuts in the complex plane, as is also the case for the simple RBM~\cite{carleo2017solving}. As a result, the usual proofs for universal representability~\cite{cybenko1989approximation,hornik1989multilayer} cannot be straightforwardly extended to these ansatze for complex parameters. 
We note that if the complex parameters were split into twice the number of real-valued parameters, causing the activation function to act independently on the real and imaginary components, the standard universal approximation theorem would apply. 
The question of universal representability for complex-valued activation functions has not been explored much except a few specific cases~\cite{voigtlaender2023universal}. Nevertheless, it is widely believed in the community that such ansatze are also universal approximators in the limit of a large number of parameters~\cite{gao2017efficient}.}, $L$ is the number of sites in the spin chain (number of units in the visible layer), $M=\alpha L$ is the number of units in the hidden layer, and $\sigma=\{\sigma_i\}$ denotes a spin-configuration on the lattice.
Note from Eq.~\eqref{eq:spin-1_RBM} that the NQS ansatz given by the spin-1 RBM is holomorphic, i.e. $\pt(\sigma)$ depends only on $\theta$, and not on $\theta^*$, i.e. $\partial\pt/\partial\te^*=0$. 

We use the above NQS \at (Eq.~\eqref{eq:spin-1_RBM}) to represent the ground states of the spin-1 BLBQ chain in different phases.
We use the infidelity measured w.r.t. the exact ground state as the loss function, for optimizing our NQS \atend.

\begin{flushleft}
{\textbf{Infidelity Optimization}}    
\end{flushleft}
For the infidelity minimization, we use the loss function given by the infidelity of the NQS w.r.t. the exact ground state $\ket{\Om}$,
    \begin{align}
        \label{eq:loss_func}\mathcal{L} = I = 1-\frac{\braket{\pt}{\Om}\braket{\Om}{\pt}}{\braket{\pt}{\pt}\braket{\Om}{\Om}},
    \end{align}
where $\theta\in\mathbb{C}^{N}$.
The loss function is minimized using the natural gradient descent (NGD) algorithm~\cite{amari1998natural,amari1998why,sorella1998green,sorella2005wave,gravina2024neural}. In the NGD method, the parameters of the \at are updated as 
\begin{align}
    \theta_{\mu} \rightarrow \theta_{\mu} -\eta \sum_{\nu}\left[\Gv+\epsilon\mathbb{1}\right]^{-1}_{\mu\nu}F_{\nu}, 
\end{align}
where $\Gv$ is the quantum geometric tensor (QGT), $\eta$ is the learning rate, $\epsilon$ is a small regularization constant which prevents the inverse from diverging, and $F_{\nu}$ is the conjugate gradient of the loss function,    
\begin{align}
    F_{\mu} = \pdd{\ml}{_{\mu}^*}=&-\frac{\braket{\Om}{\pt}}{\braket{\pt}{\pt}\braket{\Om}{\Om}}\sum_{n}\braket{n}{\Om}\pdd{\pt^*(n)}{_{\mu}^*}\nonumber\\
    &+\frac{\left|\braket{\pt}{\Om}\right|^2}{\braket{\pt}{\pt}^2\braket{\Om}{\Om}}\sum_{n}\pt(n)\pdd{\pt^*(n)}{_{\mu}^*}.
\end{align}
We compute the infidelity, the conjugate gradient, and the QGT exactly, by summing over the basis of the total $S_z=0$ sector of the Hilbert space. For better performance, we adjust the regularization constant $\epsilon$ during the course of the optimization. We also tested the infidelity minimization with a combination of ADAM~\cite{kingma2014adam} and YOGI~\cite{zaheer2018adaptive} optimizers (results not included in this article), but we found that the convergence was unstable and the solutions were significantly worse compared to those found with the current strategy. We defer an in-depth discussion on the optimization procedure to the supplementary material sec.~\ref{sec:opt-details}.

The infidelity minimization with NGD was implemented with the software library \textit{NetKet}~\cite{netket2:2019,netket3:2022,mpi4jax:2021,Sinibaldi_netket_fidelity_package_2023}.


}

\section{Supplementary Material}
\label{sec:sup_mat}

\subsection{The AKLT state}
\label{sec:AKLT}
The AKLT state, also known as a valence-bond solid~\cite{affleck1988valence}, is the ground state of the spin-1 BLBQ chain for $\theta=\arctan(1/3)$:
\begin{align}
   \label{eq:AKLT} 
    \mH_{\rm AKLT} &= \sum_{i=1}^{N-1} J \left[\Sv_i\cdot \Sv_{i+1} + \frac{1}{3}\left(\Sv_i\cdot \Sv_{i+1}\right)^2\right],
\end{align}
where $\Sv_i=(S_{ix},S_{iy},S_{iz})$ is the spin-1 operator acting on site $i$. The AKLT state is exactly known and can be represented by expressing the spin-$1$ particle by two auxiliary spin-$1/2$ particles. The spin-1 computational basis can be represented in terms of the triplet states formed with two spin-1/2 particles:
\begin{align}
    \ket{+} &= \psi_{11} = \ket{\up\up},\\
    \ket{0}  &= \psi_{12} =\frac{1}{\sqrt{2}}\left(\ket{\up\dn} + \ket{\dn\up}\right)= \psi_{21},\\
    \ket{-} &= \psi_{22} =\ket{\dn\dn},    
\end{align}
where $\{\ket{+},\ket{0},\ket{-}\}$ is the basis for the spin-1 system, and $\{\ket{\up},\ket{\dn}\}$ is the spin-1/2 basis. Then, the AKLT state is given by 
\begin{equation}    
\begin{aligned}
   \label{eq:AKLT_state}
   \ket{\Psi_{\rm AKLT}(\alpha,\;\beta)} &= \frac{1}{2^{(N-1)/2}}\;\p{}{1}\e{1}{2}\p{2}{2}\e{2}{3}\cdots\\
   &\p{i}{i}\e{i}{i+1}\cdots\p{N-1}{N-1}\e{N-1}{N}\p{N}{},
\end{aligned}
\end{equation}
where $\epsilon$ is the Levi-Civita tensor of rank 2, and repeated indices imply a summation. Note that the AKLT state Eq.~\eqref{eq:AKLT_state} is written for the case of an open boundary condition, and has two free spin-$1/2$ variables $\alpha,\;\beta$ which correspond to the two outermost spin-$1/2$s on the chain. This leads to a four-fold degeneracy of the AKLT state. In the total $S_z=0$ sector, the AKLT state is two-fold degenerate. It is interesting to note in the AKLT state that two adjacent spin-1 variables are never aligned ferromagnetically (both $+$s or both $-$s). This can be realized from Eq.~\eqref{eq:AKLT_state}, given that the Levi-Civita tensor only has off-diagonal terms. A broader consequence of this is that a $+$ ($-$) can only be followed by a $0$ or a $-$ ($+$), i.e. a state of the form $\ket{+0000000000+}$ is not allowed. Whereas a typical state could take the form $\ket{00+0-+000-0+00-0}$, which essentially has a N\'eel order when we remove all 0s. This is a consequence of a non-local order, characterized by the \textit{string order parameter} 
\begin{align}
 O_{ij} = \left\langle S_{iz}e^{i\pi\sum_{i<k<j}S_{kz}}S_{jz}\right\rangle,
\end{align} 
in the Haldane phase, which attains the maximum value at the AKLT point.\\

Furthermore, the AKLT Hamiltonian Eq.~\eqref{eq:AKLT} can be written as a sum of projection operators into the spin-$2$ subspace for every two neighboring sites, $\Po_{S=2}(i,i+1)$. The projection operator can be written by inspecting the eigenvalues of the operator $\Xo=\left(\Sv_{i}+\Sv_{i+1}\right)^2$, $S(S+1)=0,\;2,\;6$ for $S=0,\;S=1,\;S=2$ subspaces respectively. Note that the eigenspaces of the operator $\Xo$, are shared with that of the projection operators. Therefore, we can write 
\begin{align}
    \label{eq:projection_2}\Po_{S=2}(i,i+1) &= \frac{1}{24}\Xo(\Xo-2)\nonumber\\
    & = \frac{1}{2}\Sv_{i}\cdot\Sv_{i+1} + \frac{1}{6}\left(\Sv_{i}\cdot\Sv_{i+1}\right)^2+\frac{1}{3},
\end{align}
where we have used the fact that $\Xo=\Sv_{i}^2+\Sv_{i+1}^2+2\Sv_{i}\cdot\Sv_{i+1}=4+2\Sv_{i}\cdot\Sv_{i+1}$. As a consequence of Eq.~\eqref{eq:projection_2}, we can write the AKLT Hamiltonian Eq.~\eqref{eq:AKLT} as
\begin{align}
    \label{eq:projection_AKLT}\mH_{\rm AKLT} &= 2J\sum_{i=1}^{N-1}\Po_{S=2}(i,i+1)-\frac{2J}{3}(N-1)
\end{align}
An interesting observation from Eq.~\eqref{eq:projection_AKLT} is that the AKLT state has no two adjacent spin-1s living in the $S=2$ sector, so as to have the minimum energy. In the picture of the auxiliary spin-$1/2$ particles, this leads to the formation of singlet states ($S=0$) between two spin-$1/2$ particles in the adjacent sites.  

Additionally, the AKLT state can be expressed exactly as an MPS with the lowest non-trivial bond dimension $\chi = 2$~\cite{schollwock2011density}:
\begin{align}
    \label{eq:MPS-AKLT}\ket{\Psi_{\rm AKLT}(\alpha,\beta)} = \sum_{\ket{\sigma}} A_{\alpha}^{(\sigma_1)i_1}A_{i_1}^{(\sigma_2)i_2}\cdots A_{i_{N-1}}^{(\sigma_N)\beta}\ket{\sigma_1,\sigma_2,\cdots,\sigma_N},
\end{align}
where repeated indices imply a summation, $\ket{\sigma_i}$ denotes the local computational basis for the spin-$1$ particle at site $i$, with $\sigma_i=+,\;0,\;-$, and
\begin{align}
    \Av^{(+)} = \begin{bmatrix}
        0&\sqrt{\frac{2}{3}}\\
        0&0        
    \end{bmatrix},\;\;\Av^{(0)} = \begin{bmatrix}
    -\frac{1}{\sqrt{3}} & 0\\
    0 & \frac{1}{\sqrt{3}}
    \end{bmatrix},\;\;\Av^{(-)} = \begin{bmatrix}
        0&0\\
        -\sqrt{\frac{2}{3}}&0
    \end{bmatrix}.
\end{align}
$\alpha$, $\beta$ in Eq.~\eqref{eq:MPS-AKLT} are the free indices, corresponding to the two auxiliary spin-$1/2$s at the boundary for the case of an open-boundary condition. The bond dimension of the required MPS to represent the AKLT state is just one unit higher than that for the product states (which can be represented as MPSs with bond dimension $\chi=1$), making it the simplest entangled quantum state. 

\subsection{Infidelity of the NQS, and QGT, for VMC computations on the spin-1 BLBQ chain with $L=10$}
\label{sec:qgt_spectrum_vmc}

\begin{figure*}
\centering
\includegraphics[width=0.9\textwidth]{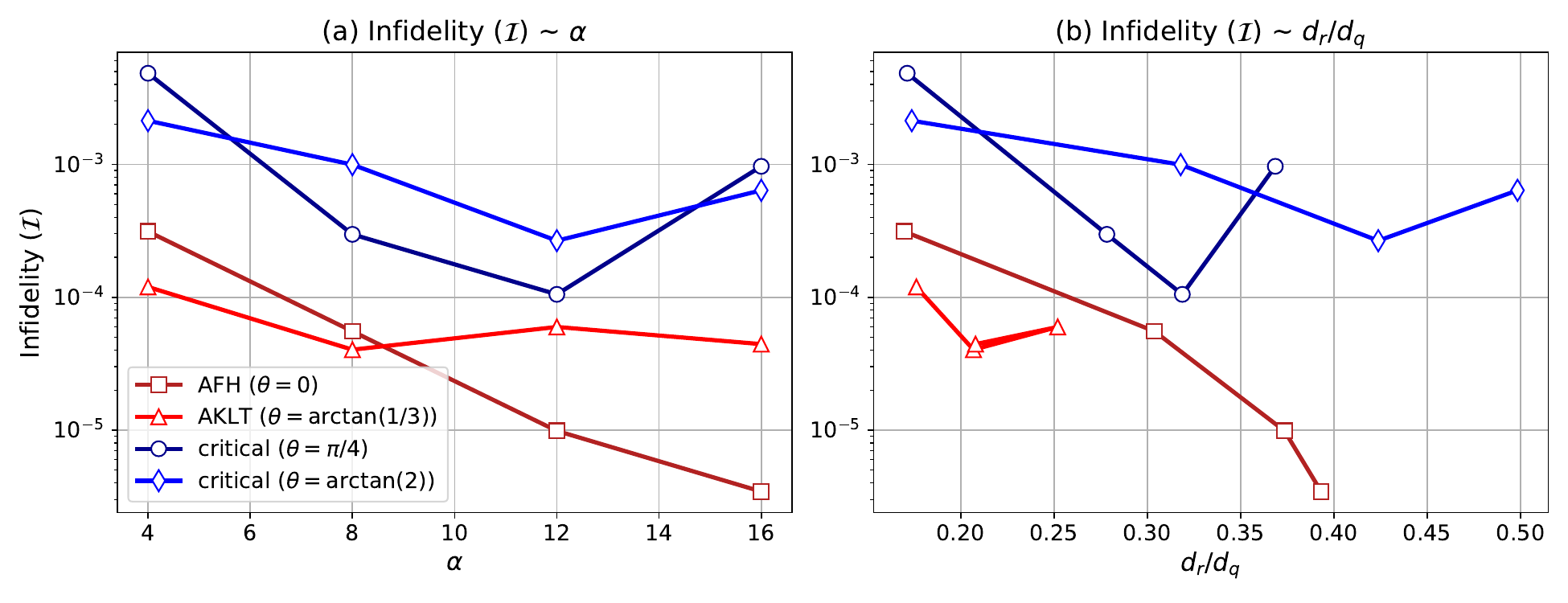}      
\caption{Infidelity and QGT for VMC on the spin-1 BLBQ chain. The infidelity ($I$), after convergence of the VMC procedure for the spin-1 BLBQ chain with length $L=10$, is shown in panel (a) as a function of $\alpha$, and in panel (b) as a  function of the ratio of the QGT rank $d_r$ (defined by a cutoff \cutoff) w.r.t. $d_q$. Results are shown for four phases of the spin-1 BLBQ chain (corresponding to the four marked points in Fig.~\ref{fig:fig_cartoon}). 
Note that the infidelity and the QGT are measured exactly at the end of the VMC optimization.} 
\label{sfig:fig_qgt_rank_vmc}
\end{figure*}

We also perform VMC optimizations~\cite{dawid2022modern} for the spin-1 bilinear-biquadratic (BLBQ) chain of length $L=10$ (with open boundary conditions), to supplement our results from the infidelity minimizations in the main text. The gradients and the QGT are estimated approximately by the Monte Carlo sampling procedure at each optimization step. However, we estimate the infidelity of the NQS w.r.t. the exact ground state (Eq.~\eqref{eq:infidelity}), and the QGT at the end of the VMC optimization exactly by a full summation over the basis states. In Fig.~\ref{sfig:fig_qgt_rank_vmc}(a), we plot the NQS infidelity, after convergence of the VMC procedure, as a function of $\alpha$. We observe that the infidelity of the NQS decreases as $\alpha$ increases, for smaller values of $\alpha$. It continues to do so up to $\alpha=16$ for the AFH phase, up to $\alpha=8$ for the AKLT phase, and up to $\alpha=12$ for the two critical phases. The infidelity for the AKLT phase saturates beyond $\alpha=8$ with small oscillations, while the NQS approximation becomes worse for the two critical phases beyond $\alpha=12$. 

Furthermore, we plot the ratio of the rank of the QGT matrix $d_r$ (defined by a cutoff \cutoff) w.r.t. the \eqd~$d_q$ in Fig.~\ref{sfig:fig_qgt_rank_vmc}(b). We observe that an initial increase in $\alpha$ leads to an increase in the ratio $d_r/d_q$, while the infidelity of the NQS decreases. This indicates a correlation between $d_r/d_q$ and the quality of the NQS approximation for smaller values of $\alpha$, as also observed for the case of infidelity minimizations (see Fig.~\ref{fig:fig_qgt_rank} in the main text). While, for larger values of $\alpha$, a larger value of $d_r/d_q$ does not necessarily lead to a better solution. This suggests that the NQS optimization is not able to utilize the additional \textit{relevant directions}, that come with increasing $\alpha$, to improve the accuracy of the NQS for larger values of $\alpha$.  

We also plot the spectra of the QGT (Eq.~\eqref{eq:QGT}), as histograms in Fig.~\ref{sfig:fig_qgt_vmc}, at the end of the VMC procedure for the four phases of the spin-1 BLBQ chain with length $L=10$, along with the cumulative distributions in the respective insets. The histograms indicate that increasing $\alpha$ leads to an increase in the number of QGT eigenvalues with moderate magnitudes. This is a bit different from the QGT spectra extracted from the converged infidelity minimizations (see Fig.~\ref{fig:fig_qgt} in the main text), where we see a significant increase in the number of large eigenvalues with an increase in $\alpha$. This may explain why the improvement of the NQS approximation eventually stops as $\alpha$ increases for the VMC optimizations, even when the ratio $d_r/d_q$ increases. 


\subsection{Details of the infidelity optimization and VMC}
\label{sec:opt-details}
\begin{figure*}
\centering
\includegraphics[width=0.9\textwidth]{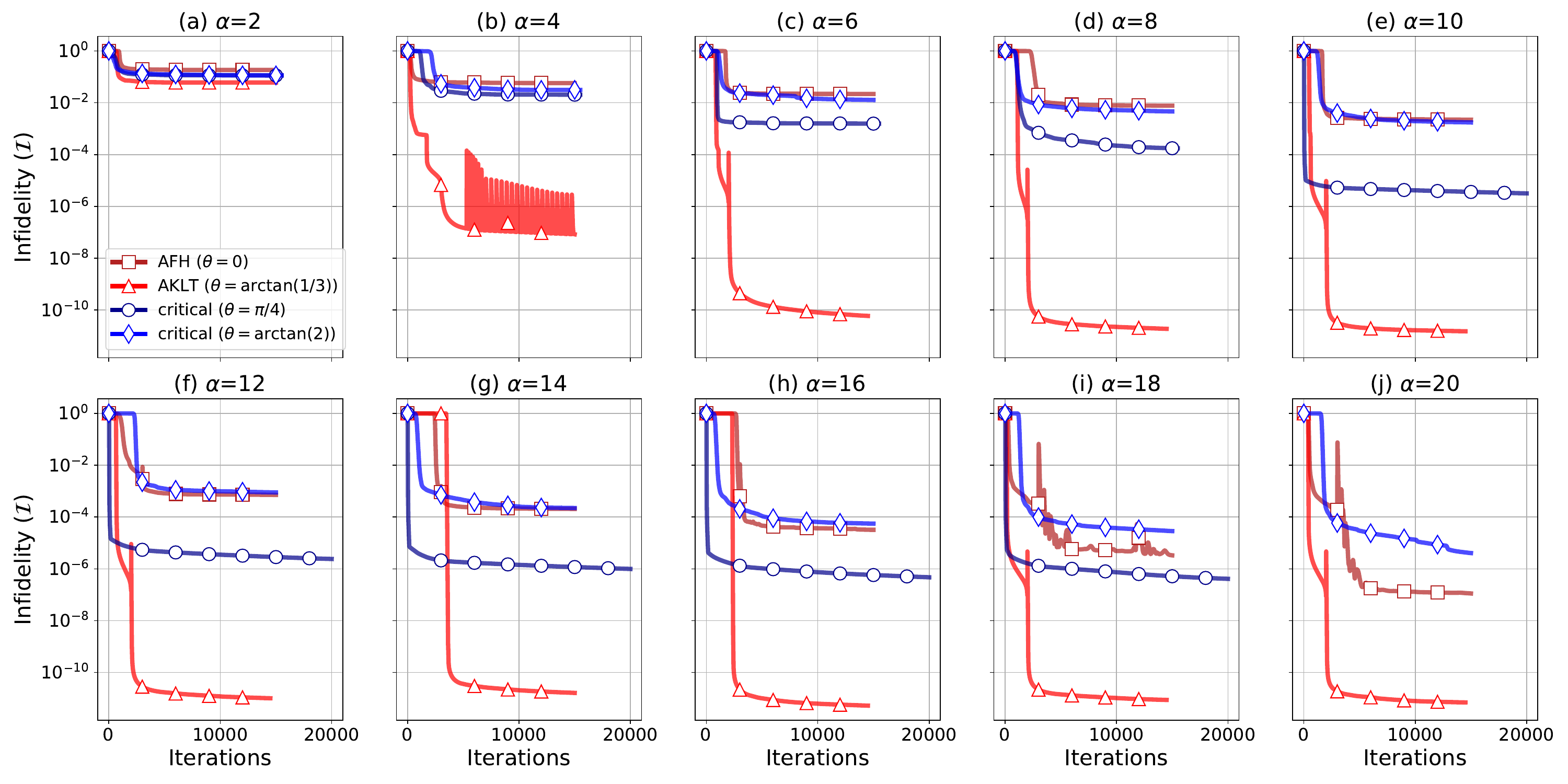}      
\caption{Evolution of the infidelity ($I$) of the NQS \at w.r.t. the exact ground state during the infidelity minimization procedure. The infidelity ($I$) is plotted (in $\log_{10}$ scale) against the minimization steps for the AFH, AKLT and the two critical phases of the spin-1 BLBQ chain with open boundary conditions for $L=10$. The panels (a)-(j) show the optimization curves for different hidden layer densities $\alpha=2,4,6,\cdots,20$ of the spin-1 RBM (Eq.~\eqref{eq:spin-1_RBM}). The markers on the curves are shown every 3000 iterations to improve readability.}
\label{sfig:fig_infid_descent}
\end{figure*}

We show the infidelity minimization curves in Fig.~\ref{sfig:fig_infid_descent}, for various phases of the spin-1 BLBQ chain with length $L=10$. The hyperparameters of the optimization are given in table~\ref{tab:details-infid}.

We impose global spin-flip symmetries on the NQS ans\"atze, as prescribed in ref.~\cite{nomura2021helping}. For the AFH ($\theta=0$) and the AKLT ($\theta = \arctan(1/3)$) states, we impose a global spin-flip symmetry with an even parity,
\begin{align}
\label{eq:parity_symm}
    \pt^{(+)}(\sigma) = \frac{1}{2}\left[\pt(\sigma) + \pt(-\sigma)\right],
\end{align}
while for the two critical phases, we impose the global spin-flip symmetry with an odd parity,
\begin{align}
    \pt^{(-)}(\sigma) = \frac{1}{2}\left[\pt(\sigma) - \pt(-\sigma)\right],
\end{align}
for lengths $L=8$, and $10$. Note that we use a global spin-flip symmetry with an even parity for all four phases for $L=12$. The correct symmetry sector for a given phase and chain length was determined by examining the exact diagonalization (ED) solution. The above symmetry prescription was used for both infidelity minimization and VMC procedures. 
\begin{table}[h!]
\begin{tabular}{|p{0.8cm}|p{1.8cm}|p{1.5cm}|p{1.8cm}|}
\hline
$L$ & Model & NQS spin-flip symmetry & Iterations  \\ \cline{1-4}
8, 10, 12 &AFH \hspace{0.1cm}($\theta=0$) & even parity & $15K$ ($L=8,10$), $20K$ ($L=12$)  \\ \cline{2-4}
 &AKLT ($\theta=\arctan(1/3)$) & even parity & $15K$ ($L=8,10$), $20K$ ($L=12$)  \\ \cline{2-4}
 &critical ($\theta=\pi/4$) & odd (even for $L=12$) & $15K$ ($L=8$), $15-20K$ ($L=10,12$)  \\ \cline{2-4}
 &critical ($\theta=\arctan(2)$) & odd (even for $L=12$)  & $15K$ ($L=8,10$), $20K$ ($L=12$) \\ \cline{1-4}
\end{tabular}
\caption{Details of the infidelity minimization procedure for the spin-1 BLBQ chain (with open boundary conditions) for lengths $L=8$, $10$, and $12$. The NQS \at is given by the spin-1 RBM Eq.~\eqref{eq:spin-1_RBM}. The values shown in the table are used for computations with all values of $\alpha$ shown in figures~\ref{fig:fig_infid},\ref{fig:fig_qgt_rank}. Note that all computations of the gradient, and the QGT were done by a full summation over the total $S_z=0$ subspace of the Hilbert space. The learning rate, and the regularization constant $\epsilon$ for the QGT matrix were chosen after a few trials for each case. We used learning rates in the range of $5\times 10^{-4}$ to $10^{-2}$, and applied a learning rate decay strategy through the optimization process. We used the regularization constant $\epsilon$ in the range of $10^{-8}$ to $10^{-6}$ for the Haldane phases, and from $10^{-6}$ to $10^{-4}$ for the critical phases, carefully selected to prevent divergence in the updates for each case. The regularization constant $\epsilon$ is also updated in the course of the optimization for improved quality of solutions~\cite{Calvetti_2000, hansen1999curve, Hansen_1993, Hansen_1992}.}
\label{tab:details-infid}
\end{table}

\begin{figure*}
\centering
\includegraphics[width=0.9\textwidth]{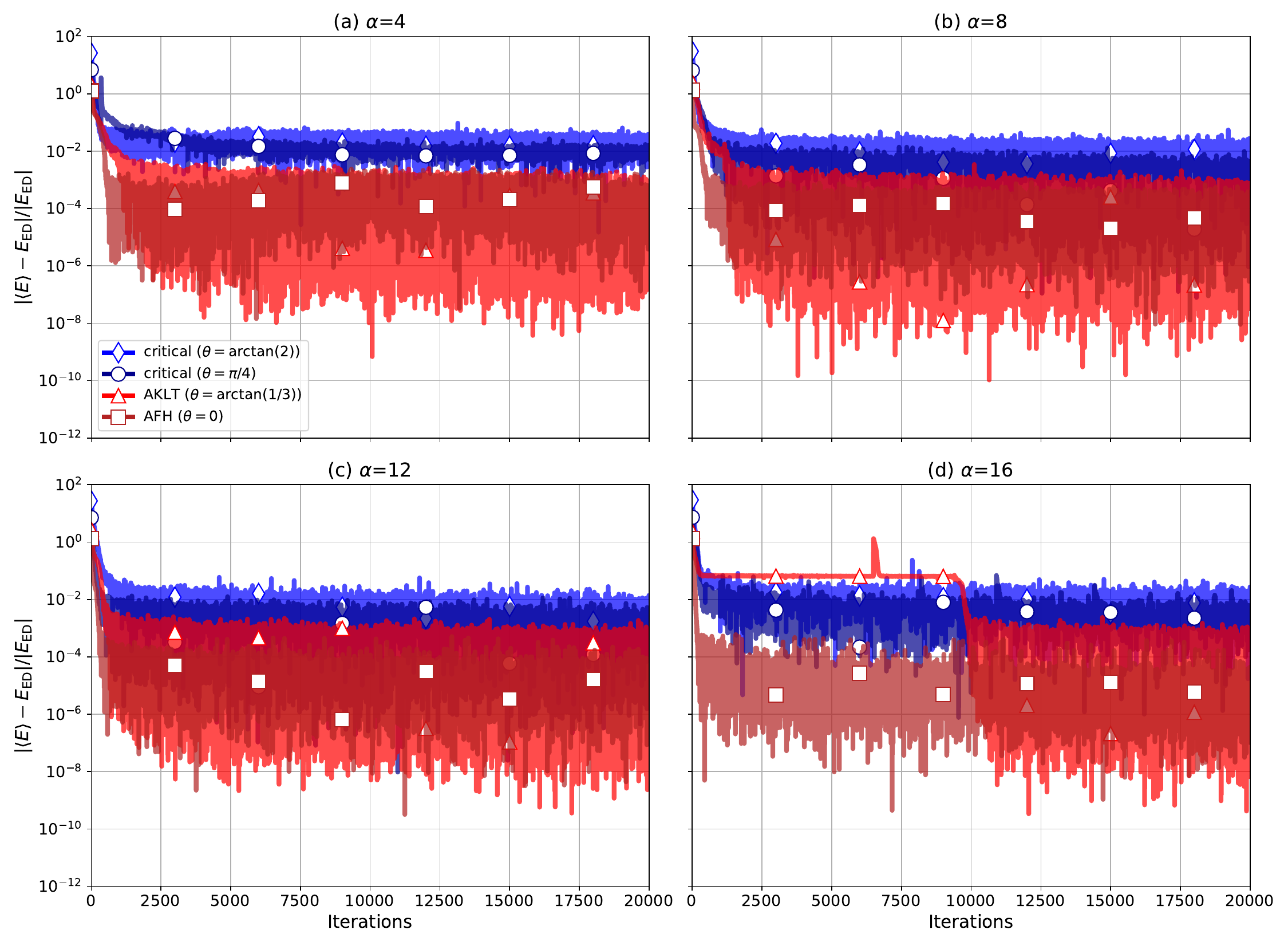}      
\caption{Evolution of the energy calculated with the NQS during the VMC procedure. The relative error in the energy calculated with the NQS w.r.t. the exact diagonalization result, i.e. $|\langle E\rangle- E_{\rm ED}|/|E_{\rm ED}|$, is plotted (in $\log_{10}$ scale) as a function of the VMC steps, for the AFH, AKLT and the two critical phases of the spin-1 BLBQ chain (with open boundary conditions) for L=10. Panels (a), (b), (c), and (d) correspond to the hidden layer densities $\alpha=4,\;8,\;12,\;{\rm and }\;16$ respectively.}
\label{sfig:fig_VMC_descent_L08}
\end{figure*}

Additionally, we also show the optimization curves for the VMC procedure for various phases of the spin-1 BLBQ chain with $L=10$ in Fig.~\ref{sfig:fig_VMC_descent_L08}(a)-(d). The hyperparameters of the computations are given in table~\ref{tab:details-VMC}. Furthermore, the number of parameters in the NQS \ats for different values of $L$ and $\alpha$ are listed in table ~\ref{tab:details-num_pars}.

\begin{table}[h!]
\begin{tabular}{|l|p{1.8cm}|p{1.6cm}|p{1.1cm}|p{1.5cm}|}
\hline
$L$ &Model & NQS spin-flip symmetry & MC samples & Iterations \\ \hline
10& AFH \hspace{0.1cm}($\theta=0$) & even parity & 4096 & $20K$ \\ \cline{2-5}
& AKLT ($\theta=\arctan(1/3)$) & even parity & 4096 & $20K$ \\ \cline{2-5}
& critical ($\theta=\pi/4$) & odd parity & 4096 & $20K$ ($45K$ for $\alpha=12$) \\ \cline{2-5}
& critical ($\theta=\arctan(2)$) & odd parity & 4096 & $20K$  \\ \hline
\end{tabular}
\caption{Details of the VMC optimization for the spin-1 BLBQ chain (with open boundary condition) with length $L=10$. The NQS \at is given by the spin-1 RBM Eq.~\eqref{eq:spin-1_RBM}. The same values are used for the optimizations with all values of $\alpha$ (unless explicitly mentioned) shown in Fig.~\ref{sfig:fig_VMC_descent_L08}. All computations were done with the symmetry constraint total $S_{z} =0$. The learning rates were in the range of $10^{-3}$ to $10^{-2}$, and we also let the learning rate decay through the optimization process. The regularization constant $\epsilon$ was taken to be between $10^{-6}$ to $10^{-4}$ for all the phases. Moreover, the regularization constant $\epsilon$ was also updated in the course of the optimization~\cite{Calvetti_2000, hansen1999curve, Hansen_1993, Hansen_1992}.}
\label{tab:details-VMC}
\end{table}

To compute the parameter update $\delta \theta$ predicted by natural gradient descent (NGD), we need to solve the linear inverse problem $\Gv \delta\theta = \bm F$. The most popular method for estimating $\delta\theta$ is the least squares method, according to which 
\begin{equation}
    \label{eqn:tikhonov_before}
    \delta \theta = \argmin_x \norm{\Gv x - \bm F}^2.
\end{equation}
The unbiased solution to this problem reads $\delta \theta = \bm \Gv^{-1} \bm F$. 
It is often the case, however, that the matrix $\Gv$ is ill-conditioned, rank-deficient, or of ill-determined rank. The computation of a meaningful approximate solution of the linear system, therefore, in general requires that the system be replaced by a nearby system that is less sensitive to perturbations. 
This replacement is referred to as \emph{regularization} \cite{Golub_1979}. 
Tikhonov regularization is one of the oldest and most popular regularization methods.
The standard formulation of Tikhonov regularization \cite{Gerth_2021} includes an $L_2$ regularization into Eq.~\eqref{eqn:tikhonov_before} which reads
\begin{equation}
    \label{eqn:tikhonov}
    \delta\theta_\epsilon = \argmin_x \norm{\Gv x - \bm F}^2 + \epsilon \norm{x}^2.
\end{equation}
This equation admits the closed-form solution 
\begin{equation}
\label{eq:preconditioning}
    \delta\theta_\epsilon = (\Gv + \epsilon\mathbb{1})^{-1} \bm F.
\end{equation}
Note that not all elements of this one-parameter family of ridge estimates $\delta\theta_\epsilon$ are viable solutions to the ubiased problem in Eq.~\eqref{eqn:tikhonov_before}. The problem of selecting the optimal regularization constant $\epsilon$ has been studied for decades, with no universal approach having been identified \cite{Bi_2022}. Classical methods such as the Morozov discrepancy principle, the L-curve method \cite{Calvetti_2000, hansen1999curve, Hansen_1993, Hansen_1992}, and generalized cross-validation \cite{Golub_1979} define somewhat heuristic criteria to select $\epsilon$. In this work we adopt the L-curve heuristic which we compute using the algorithm in Ref.~\cite{Cultrera_2020}.






\begin{table}[tb]
\centering
\vspace{1.5em}
\begin{tabular}{|p{0.8cm}|p{1.5cm}|p{1.5cm}|p{1.5cm}|p{1.2cm}|}
\hline
$L$ & $\alpha$ & number of parameters \\ \cline{1-3}
8 & 2 & 288   \\\cline{2-3}
 &4  & 560 \\\cline{2-3}
 &6  & 832 \\\cline{2-3}
 &8 & 1104 \\\cline{2-3}
 &10 & 1376 \\\cline{1-3}
 10 &2 & 440  \\\cline{2-3}
 &4  & 860 \\\cline{2-3}
 &6  & 1280 \\\cline{2-3}
 &8 & 1700 \\\cline{2-3}
 &10 & 2120 \\\cline{2-3}
 &12 & 2540 \\\cline{2-3}
 &14 & 2960 \\\cline{2-3}
 &16 & 3380 \\\cline{2-3}
 &18 & 3800 \\\cline{2-3}
 &20 & 4220 \\\cline{2-3} 
 &22 & 4640 \\\cline{1-3}
 12 &2 & 624  \\\cline{2-3}
 &4  & 1224 \\\cline{2-3}
 &6 & 1824 \\\cline{2-3}
 &8 & 2424 \\\cline{2-3}
 &10 & 3024 \\\cline{2-3}
 &12 & 3624 \\\cline{2-3}
 &14 & 4224 \\\cline{2-3}
 &16 & 4824 \\\cline{2-3} 
 &18 & 5424 \\\cline{2-3}
 &20 & 6024 \\\cline{1-3} 
\end{tabular}
\caption{The number of parameters in the spin-1 RBM \at (Eq.~\eqref{eq:spin-1_RBM}) for different sizes $L=8,10,12$ of the BLBQ chain, and different hidden layer densities $\alpha$.}
\label{tab:details-num_pars}
\end{table}

\begin{figure*}
\centering
\includegraphics[width=0.8\textwidth]{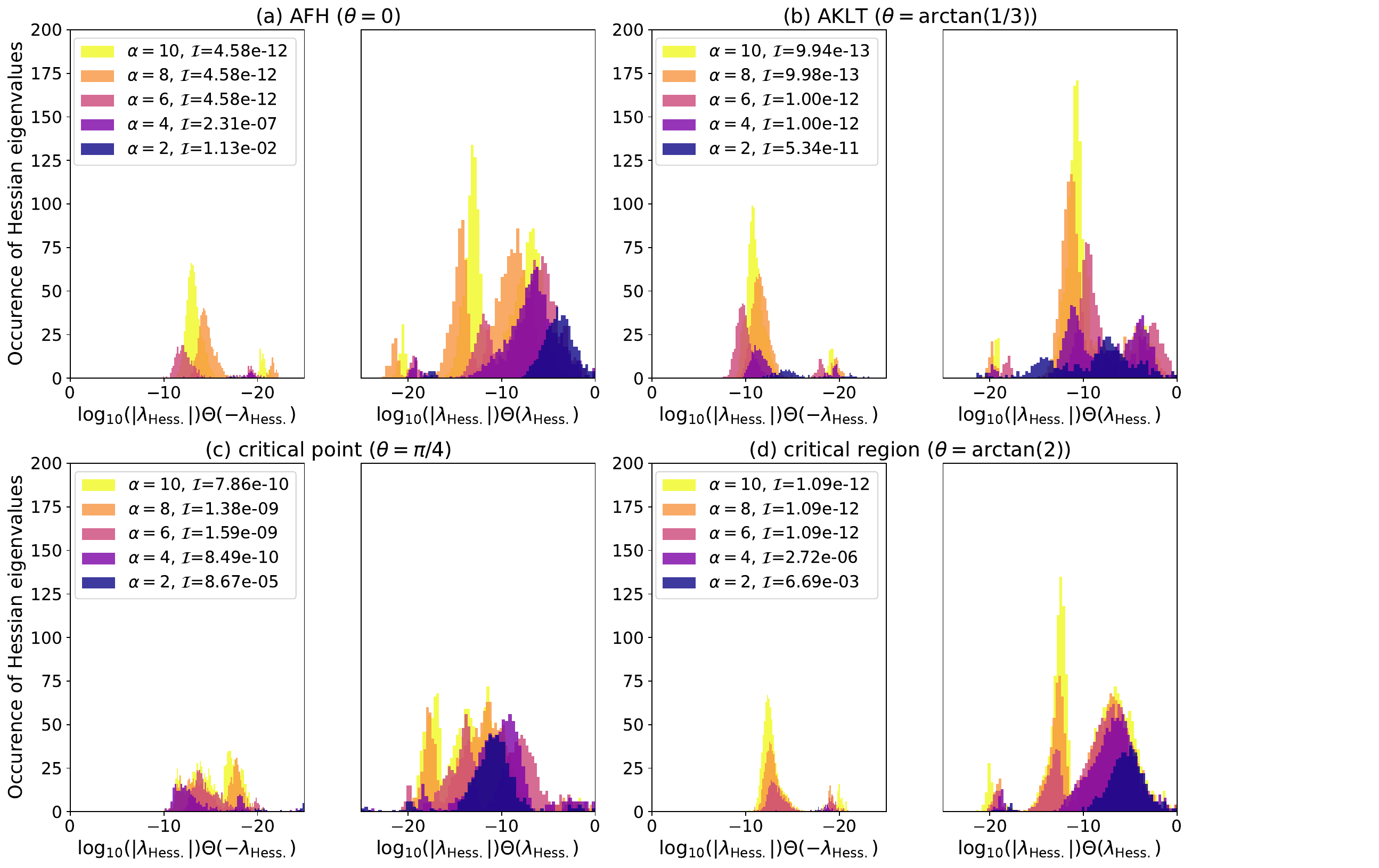}       
\caption{Spectra of the Hessian for infidelity minimization on the spin-1 BLBQ chain. The distribution of the normalized eigenvalues (w.r.t. the maximum eigenvalue) of the Hessian is shown after convergence of the infidelity minimization on the spin-1 BLBQ chain (with open boundary conditions) for $L=8$. The eigenvalues are shown in $\log_{10}$ scale. Results are shown for the AFH, AKLT, and the two critical phases of the spin-1 BLBQ chain in subplots (a), (b), (c), and (d) respectively. The left panel of each subplot shows the distribution of the negative eigenvalues (magnitudes), and the right panel of each subplot shows the distribution of the positive eigenvalues of the Hessian. In each subplot, the $x$-axis of the right panel increases from left to right, and that of the left panel increases from right to left.}
\label{sfig:fig_hess}
\end{figure*} 

\subsection{Hessian of the infidelity loss function and its relation to QGT}
\label{sec:hessian_and_QGT}
The loss function in Eq.~\eqref{eq:loss_func} is a scalar real valued function from $\mathbb{C}^{N}$ to $\mathbb{R}$, $\ml:\mathbb{C}^{N}\rightarrow\mathbb{R}$. It is straightforward to see that the loss function depends on both $\theta$ and $\theta^*$, and hence is non-holomorphic. For convenience while working with complex derivatives of non-holomorphic functions, we use the following notations~\cite{kreutz2009complex}:
\begin{align}
    \bm\theta_c = \begin{bmatrix}
        \theta\\
        \theta^*
    \end{bmatrix}\;\;,\;\;\frac{\partial}{\partial\bm\theta_c}=\begin{bmatrix}
        \pdd{}{} & \pdd{}{^*}
    \end{bmatrix}\;\;;\;\;\theta, \theta^*\in\mathbb{C}^{N}.
\end{align}

Then, we can write the complex Hessian of the loss function, following ref.~\cite{kreutz2009complex}, as
\begin{align}
\mathbb{H} &= \left(\frac{\partial}{\partial \bm\theta_c}\right)^{\dagger}\frac{\partial\ml}{\partial \bm\theta_c}\\
\label{eq:full_Hessian}    & = \begin{bmatrix}
\pdscr{}{}{\ml} & \pdscc{}{}{\ml} \vspace{0.2cm}\\
\pdsrr{}{}{\ml} & \pdsrc{}{}{\ml}
    \end{bmatrix}.
\end{align}
The generic elements of the blocks (1,1) and (1,2) in the above equation, when the loss function is given by the infidelity Eq.~\ref{eq:loss_func}, are given by
\begin{widetext}
\begin{align}
\label{}\pdscr{i}{j}{\ml} &= -\frac{\braket{\pd{i}}{\Om}\braket{\Om}{\pd{j}}}{\braket{\pt}{\pt}\braket{\Om}{\Om}}+\frac{\braket{\pt}{\Om}\braket{\Om}{\pd{j}}}{\braket{\pt}{\pt}^2\braket{\Om}{\Om}}\braket{\pd{i}}{\pt}
+\frac{\braket{\pd{i}}{\Om}\braket{\Om}{\pt}}{\braket{\pt}{\pt}^2\braket{\Om}{\Om}}\braket{\pt}{\pd{j}}\nonumber\\
&+\frac{\left|\braket{\pt}{\Om}\right|^2}{\braket{\pt}{\pt}^2\braket{\Om}{\Om}}\braket{\pd{i}}{\pd{j}}
-2\frac{\left|\braket{\pt}{\Om}\right|^2}{\braket{\pt}{\pt}^3\braket{\Om}{\Om}}\braket{\pt}{\pd{j}}\braket{\pd{i}}{\pt},\\
\pdscc{i}{j}{\ml}&= - \frac{\braket{\pds{i}{j}}{\Om}\braket{\Om}{\pt}}{\braket{\pt}{\pt}\braket{\Om}{\Om}} + \frac{\braket{\pd{j}}{\Om}\braket{\Om}{\pt}}{\braket{\pt}{\pt}^2\braket{\Om}{\Om}}\braket{\pd{i}}{\pt}+\frac{\braket{\pd{i}}{\Om}\braket{\Om}{\pt}}{\braket{\pt}{\pt}\braket{\Om}{\Om}}\braket{\pd{j}}{\pt}\nonumber\\
&+\frac{\left|\braket{\pt}{\Om}\right|^2}{\braket{\pt}{\pt}^2\braket{\Om}{\Om}}\braket{\pds{i}{j}}{\pt}
-2\frac{\left|\braket{\pt}{\Om}\right|^2}{\braket{\pt}{\pt}^3\braket{\Om}{\Om}}\braket{\pd{j}}{\pt}\braket{\pd{i}}{\pt}
\end{align}
\end{widetext}
respectively. The blocks (1,1) and (2,2) are complex conjugates of each other, and are both Hermitian. The blocks (1,2) and (2,1) are Hermitian conjugates (as well as complex conjugates) of each other. As a result, the complex Hessian matrix Eq.~\eqref{eq:full_Hessian} is Hermitian.\\

It is interesting to note that when we are at the minimum of the infidelity landscape, i.e. when $\ket{\Om}=\ket{\pt}$,  
    $$\pdscr{i}{j}{\ml} = \frac{\braket{\pd{i}}{\pd{j}}}{\braket{\pt}{\pt}} - \frac{\braket{\pd{i}}{\pt}\braket{\pt}{\pd{j}}}{\braket{\pt}{\pt}^2}= {\rm G}_{ij},$$ which is the quantum geometric tensor (QGT), and $$\pdscc{i}{j}{\ml} = \pdsrr{i}{j}{\ml} = 0.$$ In this case, the Hessian becomes
\begin{align}
    \label{eq:complex_Hessian}\left.\mathbb{H}\right|_{\ket{\Om}=\ket{\pt}} = \begin{bmatrix}
{\rm \bf G} & \mathbf{0}\vspace{0.2cm}\\
\mathbf{0}   & {\rm \bf G^*} 
    \end{bmatrix},
\end{align}
Hence, at the minimum of the infidelity landscape, the eigenvalues of the Hessian (of the infidelity loss function) are the same as that of the quantum geometric tensor (QGT), but with a degeneracy 2.  

\subsection{Spectra of the Hessian for infidelity optimization on the spin-1 BLBQ chain with $L=8$}
\label{sec:hessian_spectra}
We plot the spectra of the Hessian (Eq.~\eqref{eq:full_Hessian}) as histograms in Fig.~\ref{sfig:fig_hess}, at the end of the infidelity minimization for the four phases of the spin-1 BLBQ chain (Fig.~\ref{fig:fig_cartoon}) with length $L=8$. Eigenvalues are normalized w.r.t. the maximum eigenvalue. The positive and negative eigenvalues are shown in the right and left panels respectively in each subplot of Fig.~\ref{sfig:fig_hess}, for clarity on the nature of the landscape around the solution. 

We observe that the positive eigenvalues dominate around the converged NQSs, for all the phases that we studied. The magnitudes of the negative eigenvalues are smaller than the largest positive eigenvalue at most by a factor $\sim 10^{-6}$. This suggests that we have converged reasonably in a valley with steep positive curvatures along most directions, and only a few almost flat directions with very small negative curvatures. This holds for all values of $\alpha$ that we studied, $\alpha=2,\;4,\;6,\;8,\;{\rm and}\;10$.




\subsection{Spin-spin correlation function}
We plot the spin-spin correlation function in Fig.~\ref{sfig:fig_corr_func}, computed from the optimized NQSs for the different phases of the spin-1 BLBQ chain with length $L=10$. The correlation functions for the AFH and AKLT phases show a modulation with wave-vector $k=\pi$. This is different from that in the critical phases, where the modulation wave-vector $k\sim 2\pi/3$. Note the enhancement of the spin-spin correlation $\langle S_{0z}S_{jz}\rangle$ at the end of the chain in the AKLT state, despite the open boundary condition. This is due to the topological string order in the Haldane phase~\cite{pollmann2012symmetry}, which is strongest for the AKLT state.
\begin{figure*}[h!]
\centering
\includegraphics[width=0.9\textwidth]{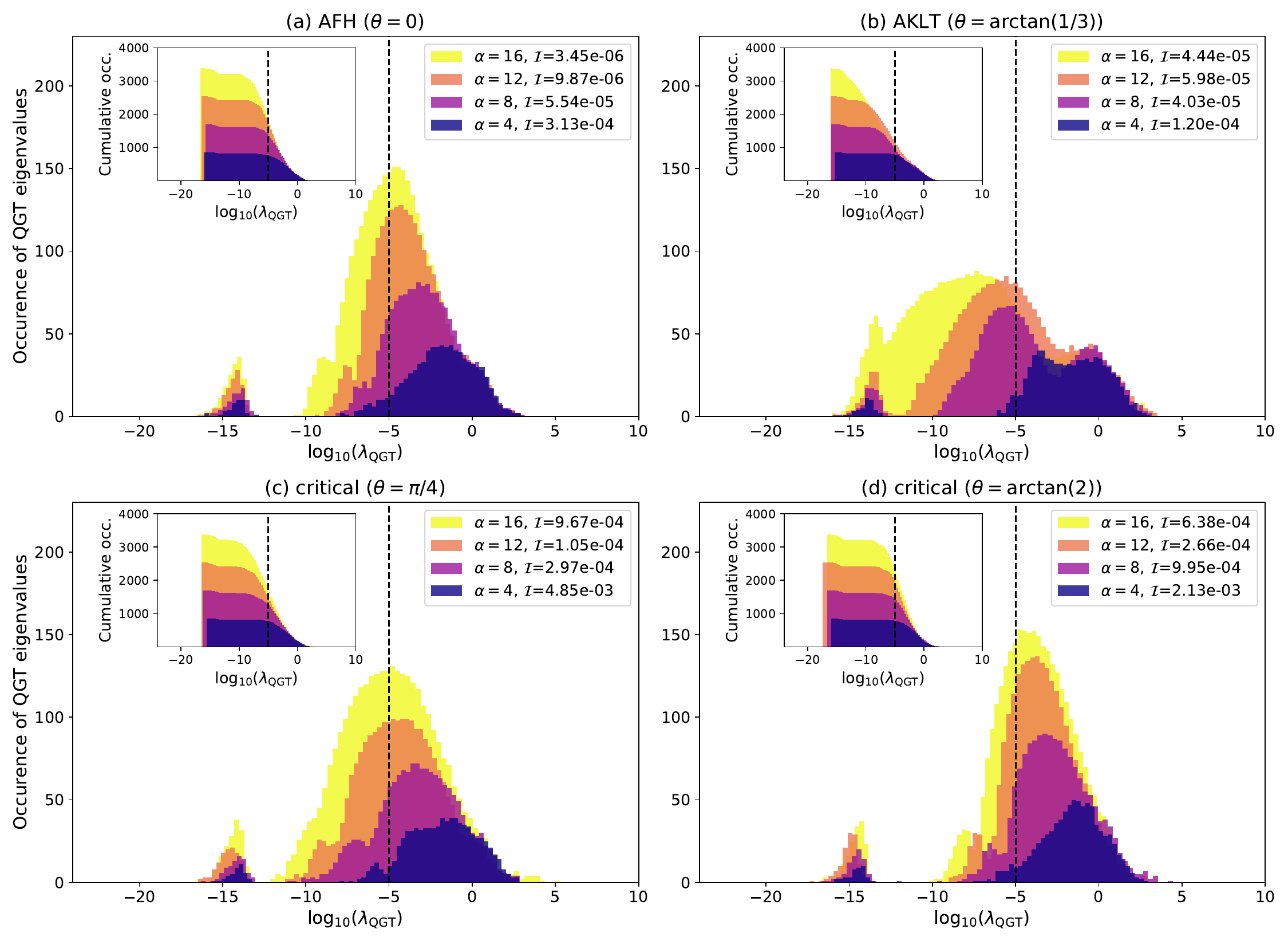}      
\caption{Spectra of the QGT for VMC on the spin-1 BLBQ chain. The distribution of the eigenvalues of the QGT is shown after convergence of the VMC optimization on the spin-1 BLBQ chain (with open boundary conditions) with length $L=10$ for the (a) AFH, (b) AKLT, and (c,d) the two critical phases in the BLBQ phase diagram (corresponding to the four marked points in Fig.~\ref{fig:fig_cartoon}). The eigenvalues are shown in $\log_{10}$ scale. Note that while the VMC procedure involves MC sampling for computing the energies, gradients, and the QGT to implement the stochastic reconfiguration method, we compute the infidelity and the QGT at the end of the optimization exactly by full summation over the total $S_z=0$ subspace of the Hilbert space. The insets
show the cumulative distribution of the eigenvalues (number of eigenvalues exceeding the value on the $x$-axis) for each case. The
dashed line marks the cutoff $10^{-5}$, used to compute the rank of the QGT $d_r$.} 
\label{sfig:fig_qgt_vmc}
\end{figure*}  

\begin{figure*}
\centering
\includegraphics[width=0.9\textwidth]{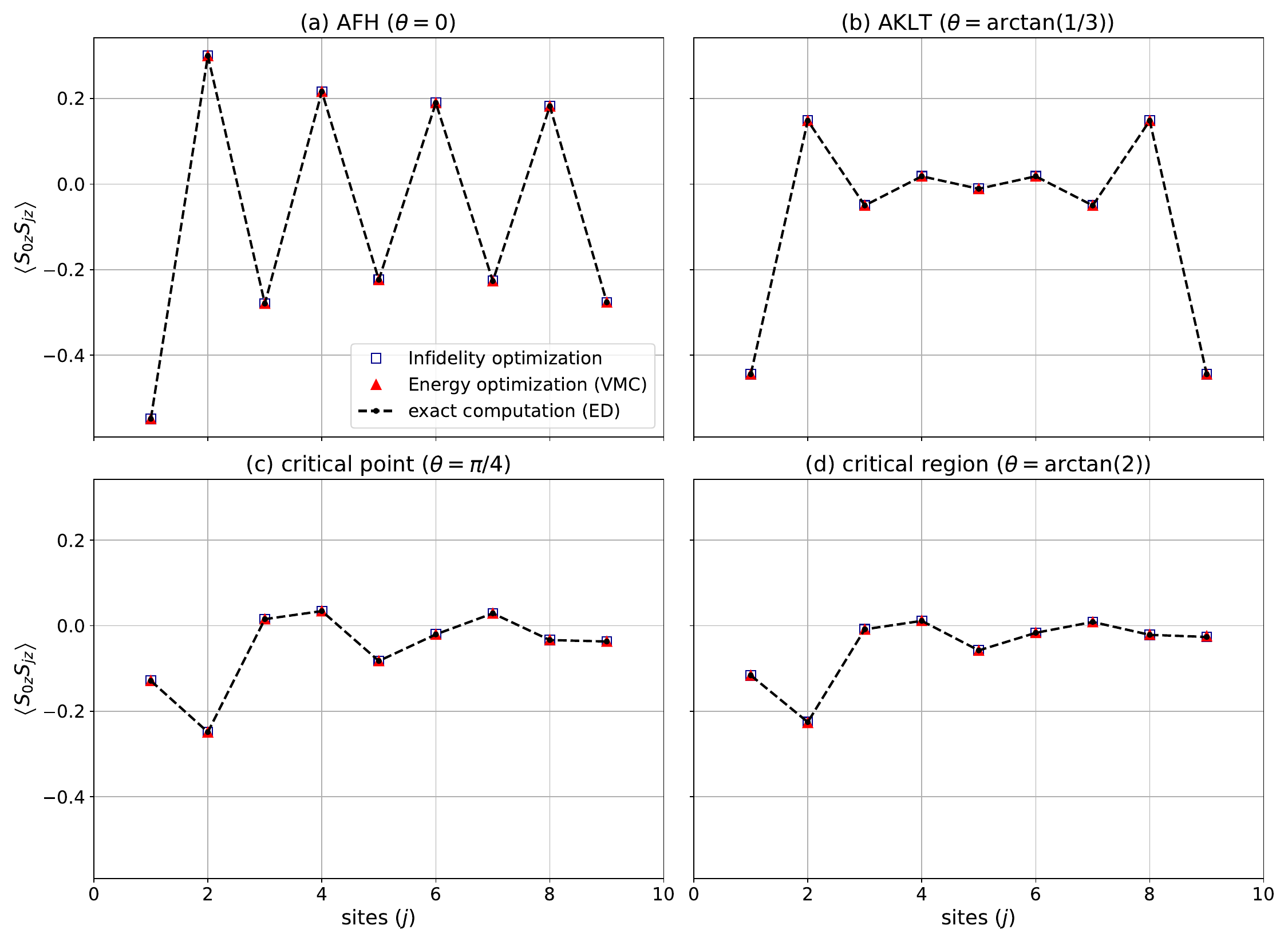}      
\caption{The real space spin-spin correlation from infidelity minimization, VMC, and exact diagonalization on the spin-1 BLBQ chain.
The spin-spin correlation function $\langle S_{0z}S_{jz}\rangle$ is shown as a function of the site $j$, for the (a) AFH, (b) AKLT, and (c, d) the two critical phases of the spin-1 BLBQ chain with length $L=10$ (with open boundary conditions). The correlation functions are computed from the optimized NQSs with hidden layer density $\alpha=16$ for the case of infidelity minimization, and VMC, and are compared with that from exact diagonalization (ED). Note that, while, the VMC procedure involves an MC sampling, we compute the correlation functions at convergence by a full summation over the total $S_z=0$ subspace of the Hilbert space.
} 
\label{sfig:fig_corr_func}
\end{figure*}

\begin{figure*}
\centering
\includegraphics[width=0.9\textwidth]{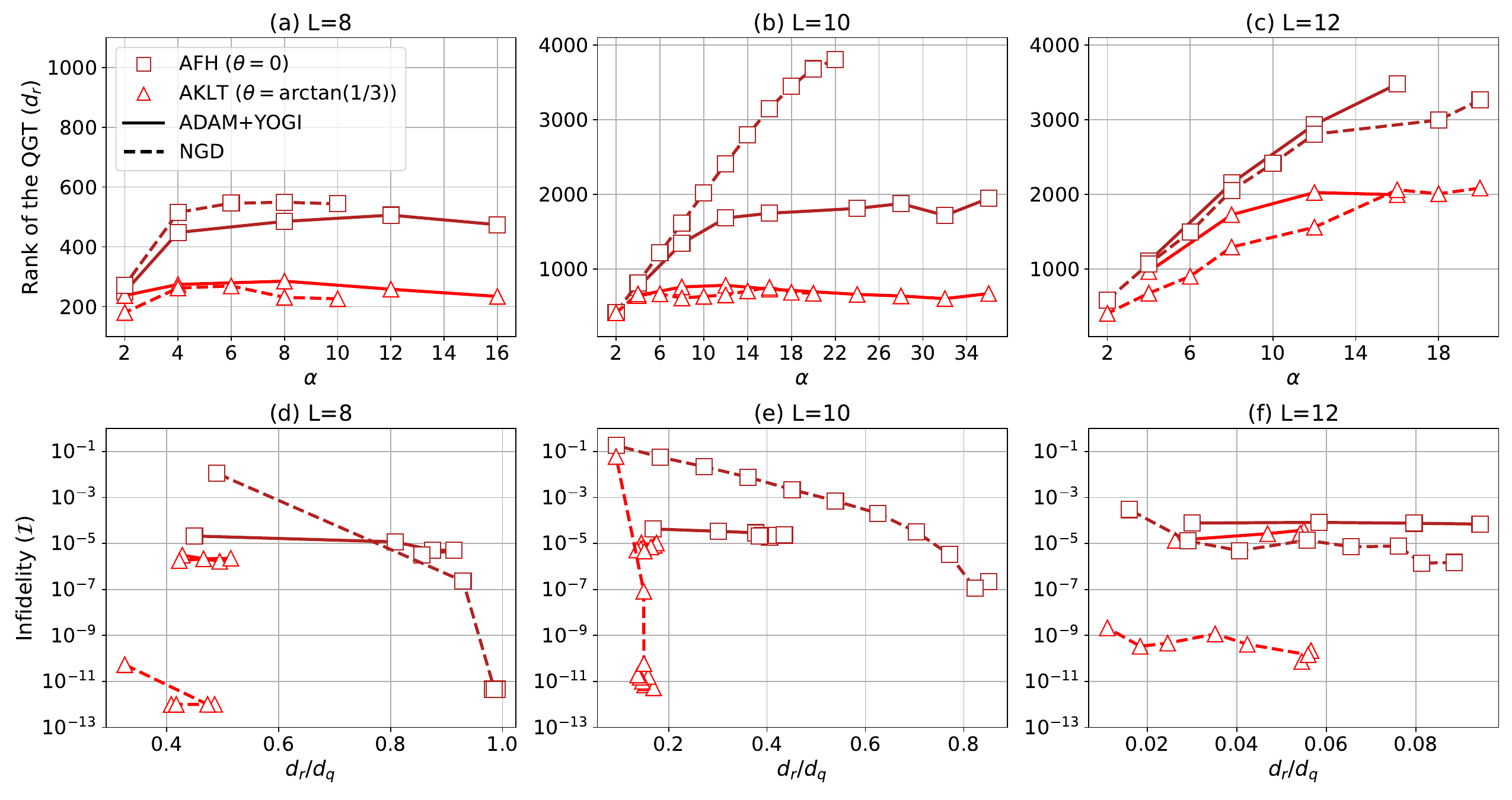}      
\caption{Performance comparison of two optimization schemes: ADAM+YOGI and natural gradient descent (NGD). (a)-(c) show the rank of the QGT as a function of the hidden layer density $\alpha$ after convergence of the infidelity minimization procedure with two different optimization schemes i) ADAM+YOGI (solid lines), and ii) NGD (dashed lines), for the AFH and the AKLT phase. (d)-(f) show the converged infidelity as a function of the ratio $d_r/d_q$ for the two optimization schemes. The rank $d_r$ is defined by the number of QGT eigenvalues greater than the cutoff $10^{-5}$ for both cases. i) For the infidelity minimization with ADAM+YOGI, first $3000$ steps were performed with the ADAM algorithm~\cite{kingma2014adam} with a learning rate $5\times 10^{-4}$, and then 117000 minimization steps were performed with the YOGI algorithm~\cite{zaheer2018adaptive} with a learning rate $3\times 10^{-4}$. The optimization was performed by an exact summation over the total $S_z=0$ subspace, and the NQS \atend, given by the spin-1 RBM (Eq.~\ref{eq:spin-1_RBM}), has the same symmetry (Eq.~\ref{eq:parity_symm}) as used in the main text. ii) The details of the NGD optimization are given in section \ref{sec:opt-details}.  
} 
\label{sfig:fig_qgt_rank_comparison}
\end{figure*} 

\subsection{Comparison of two different optimization schemes}
\label{sec:comp_optimize}
We also performed the infidelity minimization with a different algorithm for the AFH and the AKLT phases combining ADAM~\cite{kingma2014adam} and YOGI~\cite{zaheer2018adaptive} (see caption of Fig~\ref{sfig:fig_qgt_rank_comparison} for details), but the NQS approximations were generally much worse for large network widths compared to that with the natural gradient descent (NGD) optimization used in the main text. We show the rank of the QGT as a function of $\alpha$ after convergence of the infidelity minimization in Fig.~\ref{sfig:fig_qgt_rank_comparison}(a)-(c) for both optimization schemes. In addition, we show the converged infidelities as a function of the ratio $d_r/d_q$ in Fig.~\ref{sfig:fig_qgt_rank_comparison}(d)-(f) for both optimization schemes. 

We observe that the rank of the QGT saturates with $\alpha$ for the optimizations with ADAM+YOGI, as also seen for optimizations with NGD. However, in contrast to the NGD optimizations, the converged infidelity with the ADAM+YOGI scheme decreases very slowly with the rank of the QGT (or $d_r/d_q$), even increasing slowly in some cases, for example for the AKLT phase with $L=12$. The superior performance of the NGD optimization arises from preconditioning the gradient with the inverse of the QGT (Eq.~\eqref{eq:preconditioning}). This makes the local updates curvature-aware, which makes it less likely to get stuck in local minimas or saddle points. Consequently, the only limitation of the NGD comes from the rank deficiency of the QGT matrix. Therefore, we use the NGD optimization in the main text. 


\newpage

\bibliography{ref}

\end{document}